\documentclass[onecolumn,numbers]{els-mrw} 

\usepackage{amsmath,amssymb,amsfonts,amsthm,makeidx,graphicx}
\usepackage{txfonts}
\usepackage{helvet}
\usepackage{braket}
\usepackage[colorlinks=true,citecolor=blue,linkcolor=blue,urlcolor=blue]{hyperref}

\usepackage{orcidlink}

\usepackage{comment}

\usepackage{tikz}
\usepackage{tikz-feynman}
\tikzfeynmanset{compat=1.1.0}

\usepackage{subfig}

\DeclareUnicodeCharacter{2032}{\ensuremath{\prime}}
\DeclareUnicodeCharacter{2212}{\ensuremath{-}}
\DeclareUnicodeCharacter{2062}{\ensuremath{\times}}

\newcommand{\Nb}{\bar N}
\newcommand{\mpi}{M_{\pi}}
\newcommand{\Fp}{F_\pi}
\newcommand{\w}{\omega}

\begin{document}

\chapter{Hadronic parity violation: successes, challenges, and future prospects\label{HPV-chap}}

\author[1]{Susan Gardner}%
\author[2]{Jonas Karthein}%
\author[3]{Ulf-G. Mei{\ss}ner}%
\author[4]{Girish Muralidhara}
\author[5]{Petr Navratil}
\author[6]{W. Michael Snow}

\address[1]{\orgname{University of Kentucky}, \orgdiv{Department of Physics and Astronomy}, \orgaddress{Lexington, KY 40506-0055, USA}}
\address[2]{\orgname{Texas A\&M University}, \orgdiv{Cyclotron Institute, Department of Physics \& Astronomy}, \orgaddress{College Station, TX 77840, USA}}
\address[3]{\orgname{Helmholtz-Institut f\"ur Strahlen- und Kernphysik
and Bethe Center for Theoretical Physics}, \orgdiv{Universit\"at Bonn}, \\\orgaddress{D-53115 Bonn, Germany},\\ \orgname{Institute for Advanced Simulation (IAS-4)},\orgdiv{Forschungszentrum J\"ulich},\orgaddress{D-52425 J\"ulich, Germany}}
\address[4]{\orgname{University of Toronto}, \orgdiv{Department of Physics}, \orgaddress{60 St George St, Toronto, ON M5S 1A7}}
\address[5]{\orgname{TRIUMF}, \orgdiv{Theory Department}, \orgaddress{Vancouver, BC V6T 2A3, Canada}}
\address[6]{\orgname{Indiana University and IU Center for Spacetime Symmetries}, \orgdiv{Department of Physics and Center for the Exploration of Energy and Matter}, \orgaddress{Bloomington, IN 47408 USA}}

\articletag{Chapter Article tagline: update of previous edition,, reprint..}

\maketitle

\begin{abstract}[Abstract] 
Hadronic parity violation concerns the study of the 
interplay of the weak- and strong-interaction dynamics
that yields low energy, parity-violating observables in 
systems of hadrons and nuclei. 
We explain its essential features, as well as our
current understanding of its observed effects, 
describing recent theoretical and experimental progress 
in a pedagogical context. 
We provide a broad overview of ongoing research efforts 
to show how precision studies of few-nucleon systems 
can be extended to studies of complex nuclei and, ultimately, 
to new benchmarks for computations 
in the Standard Model, as well as to 
new searches for the dynamics beyond it. 
\end{abstract}

\begin{keywords}
parity violation, chiral effective theory, perturbative and lattice QCD, renormalization 
group methods, cold neutrons, nuclear anapole moments, radioactive molecules
\end{keywords} 


\section*{Key Points}
In this article we explain 
\begin{itemize}
\item that parity violation in the Standard Model 
is 
a result of the exchange of the weak 
$W^\pm$ and $Z^0$ gauge bosons between its fermions; 
\item that the strengths of the different parity-violating 
effective operators that emerge at energies 
below the gauge boson masses can be computed within 
perturbative quantum chromodynamics (QCD) to form 
an effective Hamiltonian ${\cal H_{\rm eff}}$ at scales
as low as of the order of one GeV;
\item that we could describe parity-violating 
observables in chiral effective theory in hadronic degrees of freedom 
without reference to our ${\cal H_{\rm eff}}$, although it is needed
to compute its unknown 
low-energy constants theoretically, as in lattice QCD; 
\item that hadronic parity violation has been observed
in different few-nucleon processes to show that a variety of 
low-energy constants are
important to their description, 
challenging their direct assessment from fits to the 
experimental data; 
\item that a theoretical description based on an updated assessment of 
parity-violating meson-nucleon 
couplings within QCD 
meets with 
success in describing 
parity-violating asymmetries in 
few-nucleon reactions; 
\item that 
computations of 
parity-violating observables in complex nuclei are making 
steady progress; 
\item that new experimental methods
to measure nuclear spin-dependent 
parity violation open the possibility of 
systematic studies that may yet reveal 
sources of non-Standard-Model physics
in hadronic parity violation. 
\end{itemize}

\section{Introduction
\label{sec:intro}}

The dynamics by which weak and strong physics intertwine
to yield low-energy, parity-violating observables in 
systems of hadrons and nuclei are intricate but also concrete, 
because the Standard Model (SM) 
predicts observable effects 
that have also been measured to be nonzero.
The broader study of the violation of the discrete symmetries, 
of charge-conjugation $C$ and time-reversal $T$, 
as well as of $P$, in hadrons and nuclei are a core part of the study of {\it fundamental symmetries}
in nuclear physics. That subfield comprises a chapter in this volume, of which this article is a part. 
These broader studies target observables, such as permanent electric dipole moments (EDMs), 
for which their observation at current levels of sensitivity would reflect physics beyond the SM. 
For context we note that the fundamental theory of the strong interaction, quantum chromodynamics (QCD), contains
a source of $P$ and $T$ violation, namely, $G^a_{\mu \nu} \tilde G^{a\,\mu \nu}$, that apparently does
not operate because the neutron EDM is consistent with zero\footnote{That this operator could be expected to appear 
in the QCD Lagrangian with an ${\cal O}(1)$ coupling but does not, due to the noted experimental 
constraint on the neutron EDM,  
constitutes the {\it strong CP problem}. 
}. Thus in our studies 
the presence of $P$ violation reveals the imprint of the weak interaction 
in the structure and dynamics of hadrons and nuclei. A $P$-violating observable is typically formed through the 
difference of spin-dependent cross sections to make a 
$P$-violating asymmetry, so that it is formed 
through the interference of $P$-conserving and $P$-violating amplitudes. Its 
theoretical description thus requires control over 
both $P$-conserving and $P$-violating nucleon-nucleon (NN) interactions, as
well as of the nuclear reaction dynamics and structure physics underlying the computation 
of the cross sections. 

Hadronic parity violation was first observed some ten years after the discovery of parity violation in oriented
$^{60} \vec{\rm Co}$ $\beta$-decay, 
via a nonzero circular polarization of 
$P_\gamma = (-6 \pm 1) \times 10^{-6}$  
in the gamma decay of ${}^{181}$Ta~\cite{Lobashov:1967qoh}.
Since hadronic parity violation stems from the 
interference of strong and weak interaction
effects, its observed effects are 
naturally characterized
by $G_F^{} M_\pi^2 \simeq 10^{-7}$, 
where $G_F/\sqrt{2} \simeq g^2/8 M_W^2$ is the Fermi coupling constant
and $M_\pi$ is the pion mass, and are thus 
numerically small. Heavy nuclei, through 
the appearance of closely spaced parity 
doublets, can grossly enhance the size of 
parity-violating effects~\cite{Sushkov:1982fa}, 
and many experimentally significant parity-violating effects of 
grossly larger numerical size have been 
measured~\cite{TRIPLE:1992jrn,Mitchell:1999zz}. 
Moreover, as realized shortly after the 
discovery of parity violation, a spin $1/2$ particle should also have a parity-violating coupling 
to electromagnetism --- a so-called anapole 
moment~\cite{Zeldovich1958JETP....6.1184Z}. 
With this, the search for nuclear-spin-dependent
parity violation in atomic physics was on, and, 
in 1997, a nonzero nuclear anapole moment was 
established 
in $^{133}$Cs, a nucleus of 
spin $I=7/2$~\cite{Wood1997}. 
Finding a consistent theoretical description 
of the observed parity-violating 
effects 
has proven illusive~\cite{Haxton:2013aca,Gardner:2017xyl}. 
In particular, 
reconciling the size of the 
observed parity-violating asymmetries 
in polarized $\vec{p} - p$ scattering~\cite{Eversheim:1991tg,Kistryn:1987tq} 
and 
the assessment of parity-violating effects from the 
radiative decays of excited-state 
$^{18,19}$F~\cite{Adelberger1985} with the size of the 
determined 
Cs anapole moment~\cite{Wood1997} 
has been particularly challenging~\cite{Haxton2001APV,Haxton2001PRL,Haxton2002,Haxton:2013aca}. The resolution to this puzzle may 
stem, e.g.,  from limitations in the 
meson exchange description employed~\cite{DDH1980} 
or in the treatment of nuclear structure effects~\cite{Haxton:2013aca}.  
Consequently, to separate the possibilities,  
a pivot was made: a first 
chiral effective theory description 
of the low-energy, parity-violating NN 
interaction was developed in place of the meson-exchange
description, with an intended focus on the 
study of few-nucleon observables~\cite{Zhu2005}. 
Concrete examples include the parity-violating asymmetries in 
$\vec{n}p \to d \gamma$~\cite{NPDGamma:2018vhh} 
and $\vec{n}\, {}^3{\rm He} \to p\, {}^3{\rm H}$~\cite{Gericke2020}, 
with further, pertinent cold neutron experiments possible
and under development~\cite{Gardner:2017xyl,Abele-ESS:2025fxh}. The noted cold-neutron experiments
required years, if not decades of effort, to realize the requisite
precision, and their successful conclusion~\cite{NPDGamma:2018vhh,Gericke2020}, 
and interpretation~\cite{Gardner_plb:2022mxf,Gardner_prc:2022dwi,Muralidhara:2023zcs,Viviani2014}, 
provide a starting point for a discussion not only 
of that body of work but also of the
remaining challenges and promising new directions 
offered in this overview. 

In what follows, we offer 
theoretical and experimental overviews of
hadron parity violation, noting its
theoretical description from the weak scale
to energy scales suitable to studies of 
heavy nuclei, with the associated 
experimental discussion ranging from the study of 
two-nucleon processes 
to those of complex nuclei. We conclude with a discussion
of the ongoing and upcoming synergies between 
theory and experiment before offering a final
summary. 

\section{Theoretical Overview}\label{sec:theory}

\begin{figure}
\includegraphics[width=12cm]{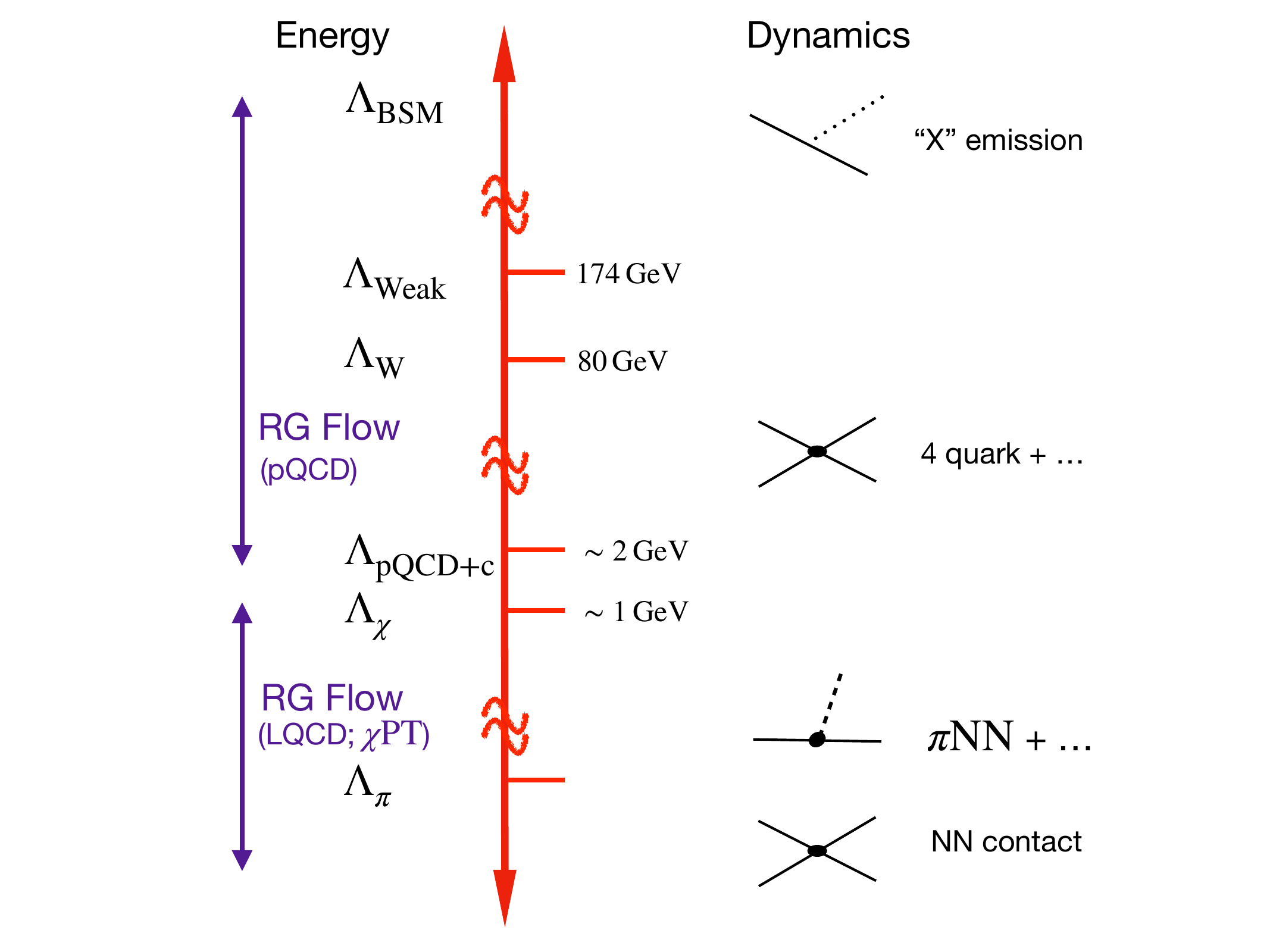}
    \caption{
    A schematic energy ladder for 
    the theoretical description of hadronic parity violation, showing how its 
    dynamical content evolves 
    from very high energy  scales, 
    just above some $\Lambda_{\rm BSM}$ 
    where a new particle
    ``$X$'' 
    could be produced directly, and 
    putatively well above the scale of electroweak symmetry breaking in the SM, 
    $\Lambda_{\rm Weak}=(2\sqrt{2}G_F)^{-1/2}  = v/\sqrt{2}$, with 
    $v =246\,\rm GeV$, 
        to the very low energy scales pertinent to the study of hadronic parity violation in nuclei. The double jagged lines denote breaks in the energy scale. The RG
        controls the theoretical evolution from high energies to low energies, with the evolution in perturbative QCD (pQCD) ending near 2 GeV. Upon the transition to chiral effective theory for scales below about 1 GeV, the evolution to still lower energies is nonperturbative in nature. At the very lowest     energies, we note that the 
    $\pi$ meson would no longer play a dynamical role in the theory, though the splay of energy scales required to describe finite nuclei may make 
    this last step of more limited applicability. 
We note different paths to the computation of the
parity-violating $\pi$-N coupling 
constant, $h_\pi^1$, at a scale of
$1-2\, \rm GeV$ at the end of 
Sec.~2.2 --- and we refer to the text for all details. 
    }
    \label{fig:HPVscales}
    \end{figure}

Parity is not a symmetry of the SM, and this follows 
from the chiral, or handed, 
nature of the weak gauge boson couplings to fermions. 
Possibly, too,  physics beyond the SM  
or new particles, 
produced at energies beyond the weak scale, e.g., also appear
and modify the expected strength of any parity-violating 
observables --- with this possibility to be limited by comparison
with experiment. 
An essential problem is thus to determine how the effective 
interactions, and their 
couplings, that emerge just below $\Lambda_{\rm W}$, the $W$ mass scale,
evolve to the energy scales pertinent to 
high-precision, experimental tests of
parity violation in hadrons and ultimately in nuclei.
We provide a schematic of the needed
evolution in Fig.~\ref{fig:HPVscales}, with 
the description changing from quark to hadron
degrees of freedom at about 1 GeV. 

To explain this, we recall that a physical coupling 
is controlled by the energy scale of the 
experiment that would determine it, 
and the evolution we note follows from 
the computed change in that coupling with energy scale, 
where we note~\cite{Zee:2010qce} for an illuminating overview.
The evolution we describe 
is essential to connecting an effective operator in the SM 
at a scale just below $\Lambda_{\rm W}$ to that of 
experiments. It arises overwhelmingly from 
QCD radiation 
from that operator's quarks, with that 
process also generating additional operators, 
ultimately forming a complete set 
under the considered radiative corrections. 
This set of operators combine to form 
the effective Hamiltonian for hadronic 
parity violation, and its 
evolution from a renormalization 
scale $\mu$ to another scale 
is determined with so-called renormalization 
group (RG) methods, using perturbative QCD~\cite{Buras:1998raa,Buchalla_1996}. 
Such methods can be used to evolve downward
in $\mu$, 
to regions of larger
strong coupling, as long as perturbative QCD remains valid. This can be done 
down to a scale of 
$\Lambda_{\rm pQCD+c}\sim 2\, \rm GeV$,
at which the charm quark is still an active 
(or open) 
degree of freedom. At lower energies, 
below about $\Lambda_{\chi} \sim 1\, \rm GeV$, 
our interactions would be formulated in 
chiral effective theory in hadron degrees of freedom. 
The apparent gap in the two descriptions 
stems, as shown in Fig.~\ref{fig:HPVscales}, is an artifact of the shift from the
quark to hadron descriptions and is not an intrinsic 
limitation. If, rather, we were to work in a theoretical 
description that could operate 
through the $\sim 1-2\, \rm GeV$ energy regime, then
there would be no gap. 
An concrete example is given 
by the analysis, e.g., of nonleptonic 
kaon decays in dual QCD~\cite{Bardeen:1986vz,Bijnens_deltaI_1/2:1998ee,Buras_lowscaleevol:2014maa}. 
Dual QCD is a model stemming from  
the dual representation of QCD for a large
number of colors $N_{\rm c}$
as a theory of weakly interacting
mesons~\cite{tHooft:1973alw,tHooft:1974pnl,Witten:1979kh},
and it  has been used to analyze 
the anatomy 
of the $|\Delta I|=1/2$ rule in $K\to \pi\pi$ decay~\cite{Bardeen:1986vz,Bijnens_deltaI_1/2:1998ee,Buras_lowscaleevol:2014maa}. 
The ``rule'' speaks to the large enhancement of the isospin $I=0$ 
$\pi\pi$ final-state amplitude over that of the $I=2$ combination, and we 
note \cite{Bijnens_deltaI_1/2:1998ee} 
for an analysis in a similar spirit. 
Concretely, it is found that the quark-gluon RG
evolution is slow over large scales, say from 
$M_{\rm W}$ to $\sim 1\,\rm GeV$ 
and fast in meson degrees of freedom at smaller 
scales, say from below $1\,\rm GeV$ to zero~\cite{Buras_lowscaleevol:2014maa}.
Thus we anticipate that it will be important to 
analyze the RG
flow at scales
below $1\, \rm GeV$ carefully. Nevertheless, 
we can glean a sense of those expected refinements through 
evaluations of the parity-violating pion-nucleon 
coupling constant $h_\pi^1$ on either side of the 
``gap'' at $\sim 1-2\, \rm GeV$, as we shall detail, and thus we regard 
such assessments with high interest. 
Certainly, in this energy regime computations of SM 
dynamics are still being clarified and tested. 

In remainder of this section, we offer an overview of the ideas that drive the description of 
hadronic parity violation, as well as of how
that description changes from high to low energy scales, 
noting ongoing challenges. 

\subsection{Parity violation in the SM  
\label{sec:th-SM}}

Parity symmetry corresponds to invariance under spatial inversion. In a field-theoretic setting, the parity operation inverts the spatial coordinates of a field,
effectively changing the ``handedness" in space. 
The parity operation in the SM 
is defined through its action on fermion field $\psi(x)$: 
$   P \psi(t,\mathbf{x}) P^{-1} = \eta_P \gamma^0 \psi(t,-\mathbf{x})$,
where $\eta_P$ is the phase factor of parity transformation. For Dirac fermion-fields, this is conventionally chosen to be $-1$.   
Parity conservation/violation of the terms in the Lagrangian of the SM, 
$\mathcal{L}_{SM}$, is determined by the transformation properties of the constituent Dirac bilinears. In Table \ref{parity table} the action of parity on different Dirac bilinears are listed. 

\begin{table}[htp]
    \centering
    \renewcommand{\arraystretch}{1.50}
    \begin{tabular*}{\textwidth}{@{\extracolsep{\fill}} l c c c}
    \hline  
         & Dirac bilinear & P-operation & Parity \\ 
    \hline  
    Scalar (S)
        & $\Bar{\psi}\psi(x)$ 
        & $\Bar{\psi}\psi(Px)$ 
        & +1 \\   
    Pseudoscalar (P)
        & $\Bar{\psi}\gamma^5\psi(x)$
        & $-\Bar{\psi}\gamma^5\psi(Px)$ 
        & $-1$ \\ 
    Vector (V)
        & $\Bar{\psi}\gamma^{\mu}\psi(x)$ 
        & $(-1)^{\mu}\Bar{\psi}\gamma^{\mu}\psi(Px)$ 
        & $(-1)^{\mu}$ \\ 
    Axial-vector (A)
        & $\Bar{\psi}\gamma^{\mu}\gamma^{5}\psi(x)$ 
        & $-(-1)^{\mu}\Bar{\psi}\gamma^{\mu}\gamma^{5}\psi(Px)$
        & $-(-1)^{\mu}$ \\ 
    \hline 
    \end{tabular*}
    \caption{The result of parity operation on Dirac bilinears ~\cite{Peskin:1995ev}. (Convention: $(-1)^{\mu} = 1$ for $\mu=0$ and $(-1)^{\mu} = -1$ for $\mu=1,2,3$.)}
    \label{parity table}
\end{table}

Parity violation arises naturally from the chiral structure of the weak interaction, which couples exclusively to left-handed fermions and right-handed antifermions. To illustrate this, consider a term analogous to the effective Hamiltonian of Fermi theory (it is convenient to introduce the shorthand notation  $\bar{\psi}_1 \gamma^{\mu}(1-\gamma^5)\psi_2 \equiv (\bar{\psi}_1 \psi_2)_{V-A}$),
\begin{equation}
    \begin{split}
    P\,\mathcal{L}_{\mathrm{Left}}\,P^{-1}
    &= P\,\frac{G_F}{\sqrt{2}}
    \left(\bar{\psi}_1\psi_2\right)_{V-A}
    \left(\bar{\psi}_3 \psi_4\right)_{V-A}
    P^{-1} \\
    &= \mathcal{L}_{\mathrm{Right}},
    \end{split}
\end{equation}
where $\mathcal{L}_{\mathrm{Right}}$ denotes the interaction obtained by replacing each left-handed chiral projector $(1-\gamma^5)$ with its right-handed counterpart $(1+\gamma^5)$. This demonstrates explicitly that the left-handed weak interaction is not invariant under parity. In particular, parity violation originates from the presence of mixed vector--axial-vector ($VA$) terms in the interaction. This built-in asymmetry has far-reaching consequences for particle phenomenology and the structure of the theory itself.\\

\noindent \underline{Hadronic Parity Violation}: \\

\begin{figure}[h!]
    \centering
    \scalebox{0.75}{
\begin{tikzpicture}
  \begin{feynman}
    \vertex (a2);
    \node[above right=1.5cm of a2] (a3);
    \vertex[above left=1.5cm of a2] (a1);
    \vertex[above right=1cm of a2] (g1);
    \vertex[above left=1cm of a2] (g3);
    \vertex[below=1.5cm of a2] (b2);
    \vertex[below right=1.5cm of b2] (b3);
    \vertex[below left=1.5cm of b2] (b1);
    \vertex[below right=1cm of b2] (f1);
    \vertex[below left=1cm of b2] (f3);
    \diagram* {
      {[edges=fermion]
        (b1) -- (b2) -- (b3),
        },
      (a1) -- [fermion] (a2) -- [fermion](a3),
      (a2) -- [boson, edge label'=W/Z] (b2),
      (g1) -- [gluon, edge label'= g](g3)
    };
  \end{feynman}
\end{tikzpicture}}
    \quad
\scalebox{0.75}{
    \begin{tikzpicture}
  \begin{feynman}
    \vertex(a2);
    \node[above right=1.5cm of a2] (a3);
    \vertex[above left=1.5cm of a2] (a1);
    \vertex[above right=1cm of a2] (g1);
    \vertex[above left=1cm of a2] (g3);
    \vertex[below=1.5cm of a2] (b2);
    \vertex[below right=1.5cm of b2] (b3);
    \vertex[below left=1.5cm of b2] (b1);
    \vertex[below right=1cm of b2] (f1);
    \vertex[below left=1cm of b2] (f3);
    \diagram* {
      {[edges=fermion]
        (b1) -- (b2) -- (b3),
        },
      (a1) -- [fermion] (a2) -- [fermion](a3),
      (a2) -- [boson, edge label'=W$^{\pm}$/Z] (b2),
      (g1) -- [gluon, edge label'= g](f1)
    };
  \end{feynman}
\end{tikzpicture}}
    \quad
\scalebox{0.75}{
    \begin{tikzpicture}
  \begin{feynman}
    \vertex(a2);
    \node[above right=1.5cm of a2] (a3);
    \vertex[above left=1.5cm of a2] (a1);
    \vertex[above right=1cm of a2] (g1);
    \vertex[above left=1cm of a2] (g3);
    \vertex[below=1.5cm of a2] (b2);
    \vertex[below right=1.5cm of b2] (b3);
    \vertex[below left=1.5cm of b2] (b1);
    \vertex[below right=1cm of b2] (f1);
    \vertex[below left=1cm of b2] (f3);
    \diagram* {
      {[edges=fermion]
        (b1) -- (b2) -- (b3),
        },
      (a1) -- [fermion] (a2) -- [fermion](a3),
      (a2) -- [boson, edge label'=W$^{\pm}$/Z] (b2),
      (g1) -- [gluon, edge label'= g, half right](f3)
    };
  \end{feynman}
\end{tikzpicture}}
\quad
    \scalebox{0.75}{
    \begin{tikzpicture}
  \begin{feynman}
    \vertex (a1){\(\psi_2\)};
    \vertex[below right=of a1] (a2);
    \vertex[above right=1cm of a2] (a3){\(\psi_1\)};
    \vertex[below=1cm of a2] (a4);
    \vertex[below=1cm of a4] (a5);
    \vertex[below=1cm of a5] (a6);
    \vertex[right=1cm of a6] (a7){\(q\)};
    \vertex[left=1cm of a6] (a8){\(q\)};
    \diagram* {
      {[edges=fermion]
        (a8) -- (a6) -- (a7),
        },
      (a1) -- [fermion] (a2) -- [fermion](a3),
      (a2) -- [boson, edge label'=W$^{\pm}$/Z] (a4),
      (a5) -- [gluon, edge label'=g] (a6),
      (a4) -- [fermion, half right] (a5),
      (a5) -- [fermion, half right] (a4),
    };
  \end{feynman}
\end{tikzpicture}}
    \quad
    \scalebox{0.75}{
    \begin{tikzpicture}
  \begin{feynman}
    \vertex (a1){\(\psi_2\)};
    \vertex[below right=of a1] (a2);
    \vertex[right=0.5cm of a2] (a4);
    \vertex[above right=1cm of a4] (a3){\(\psi_1\)};
    \vertex[right=0.25cm of a2](b);
    \vertex[below=1cm of b] (a5);
    \vertex[below=1.4cm of a5] (a6);
    \vertex[right=1cm of a6] (a7){\(q\)};
    \vertex[left=1cm of a6] (a8){\(q\)};

    \diagram* {
      {[edges=fermion]
        (a8) -- (a6) -- (a7),
        },
      (a1) -- [fermion] (a2),
      (a4) -- [fermion] (a3),
      (a5) -- [gluon, edge label'=g] (a6),
      (a2) -- [boson, edge label'=W$^{\pm}$/Z] (a4),
      (a2) -- [fermion, half right] (a5),
      (a5) -- [fermion, half right] (a4),
    };
  \end{feynman}
\end{tikzpicture}
    }
    \caption{Leading order QCD corrections to 2 $\to$ 2 weak processes. Curly lines represent gluon dressings  to tree-level W$^\pm$/Z$^0$ boson mediated weak interactions, which are represented by wavy lines.}
\label{LO qcd corrections}
\end{figure}
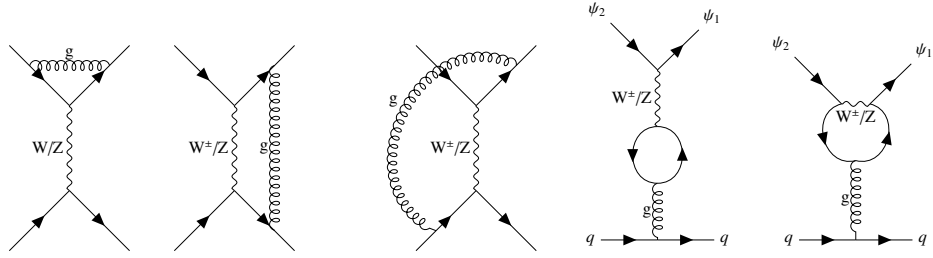

Baryons and mesons, through their underlying quark content,
participate in weak interactions that exhibit parity-violating
effects, collectively
referred to as hadronic parity violation.
This includes low-energy weak interactions involving nucleons and
nuclei. Such processes are experimentally accessible through
observables such as spin--momentum
correlations and circular polarization of emitted photons.
However, first-principles theoretical calculations of these observables remain challenging. The difficulty arises from the subtle interplay between electroweak and strong interactions, as weak interactions among nucleons and constituent quarks receive significant corrections from QCD gluon loops. (See Fig.~\ref{LO qcd corrections}).
These issues are compounded by QCD renormalization-group evolution effects and the inherently nonperturbative nature of the strong interaction at hadronic energy scales.

To address these challenges, Desplanques, Donoghue, and Holstein (DDH)~\cite{DDH1980} introduced a phenomenological NN 
interaction model based on single-meson exchange to describe hadronic parity violation. In this framework, parity-violating weak interactions are parameterized in terms of effective meson-nucleon couplings. Parity-violating observables in nuclear processes are then expressed directly in terms of these couplings.
The DDH model further provides estimates for the values of the 
meson-nucleon couplings 
by incorporating QCD renormalization effects through phenomenological enhancement factors, thereby offering a practical bridge between the underlying weak interaction at the quark level and parity-violating observables at hadronic scales. 
However, increasingly precise experimental measurements of hadronic parity violation in few-body nuclear systems are now underway. The NPDGamma collaboration~\cite{NPDGamma:2018vhh} measures the parity-violating asymmetry arising from neutron spin reversal in the process $\vec n + p \to d + \gamma$, providing a determination of the isovector pion-nucleon weak coupling. Similarly, the $n ^3$He collaboration~\cite{Gericke2020} constrains a linear combination of vector-meson-nucleon weak couplings through measurements of parity violation in the reaction $\vec n + {}^3\mathrm{He} \to t + p$. These advances underscore the need for a rigorous formulation of hadronic parity violation within an effective-field-theory (EFT) 
framework.
\\

\noindent \underline{Effective Field Theory Framework for Weak Interaction Studies}\\

Consider the classic example of 
the theory of 
$\beta$ decay 
with the $V-A$ law~\cite{Fermi:1934hr,FeynmanGellmann1958,Sudarshan:1958vf}.
The effective Hamiltonian 
describing the conversion of neutron to a proton with the 
associated emission of an electron and an anti-neutrino
is:
\begin{equation}
    \mathcal{H}_{\mathrm{eff}}
    = \frac{G_F}{\sqrt{2}} \cos\theta_c\,
    (\bar{u}d)_{V-A}
    (\bar{e}\nu_e)_{V-A},
\end{equation}
where $\theta_c$ is the Cabibbo mixing angle describing two-flavor
quark mixing, and $G_F$ is the Fermi coupling constant. 
The rest is the semi-leptonic operator 
needed to give 
the physical process. Drawing inspiration from this, a first step in constructing an effective description of hadronic weak decay processes is the formulation of an effective Hamiltonian using the operator product expansion~\cite{Buras:1998raa}. When a system is probed at energies below a characteristic scale $\Lambda$, the effects of heavy degrees of freedom decouple from the low-energy dynamics. 
For weak processes at low energies, the heavy electroweak gauge bosons are integrated out, since the $W$ boson mass is $M_W \simeq 80~\mathrm{GeV}$ (with the $Z$ boson being heavier). 
The resulting effective theory of weak interactions 
between $\sim 1~\mathrm{GeV}$ and $M_W$ 
can be described by supplementing the renormalizable QCD Lagrangian (with QED included if the desired precision requires it)
by a tower of higher-dimensional, non-renormalizable operators of (mass) dimension $4+n$, suppressed by appropriate powers of $\Lambda$.
If our attention is restricted to 
processes in which the change in baryon number B matches the change in lepton number L, the dimension-five operators are absent~\cite{Kobach:2016ami},
and the effective Hamiltonian consists of dimension-six four-fermion operators.
The coefficients of these operators can be determined by matching the outcomes of 
the full theory and the effective theory
at a chosen matching scale.

The effective Hamiltonian for weak processes as an operator product expansion, i.e., as 
a series of operators and Wilson coefficients is
\begin{equation}
        \mathcal{H}_{\mathrm{eff}} = \frac{G_F}{\sqrt{2}}\sum_i V_{\mathrm{CKM}}^i C_i(\mu) \Theta_i (\,\mu) \,
\end{equation}
where $\mathrm{V_{CKM}}$ are the Cabibbo-Kobayashi-Maskawa factors \cite{Kobayashi:1973fv}. The 
operators ($\Theta_i(\mu)$) are the effective vertices of the interactions, and the Wilson coefficients 
($C_i(\mu)$)
are the effective couplings. The interplay between the electroweak and the strong interactions are captured by the appearance of non-trivial effective vertices and modifications to the strengths of the effective couplings at lower energy scales.

Although both the operators and their associated Wilson coefficients depend on the renormalization scale $\mu$, physical amplitudes and observables computed from the effective Hamiltonian must be independent of this arbitrary scale choice. The Wilson coefficients $C_i(\mu)$ encode contributions from short-distance physics at scales above $\mu$, while the hadronic matrix elements $\langle \Theta_i(\mu) \rangle$ contain information from long-distance dynamics below $\mu$. Scale-independent physical predictions imply  that the $\mu$ dependence of the Wilson coefficients cancel that of the matrix elements~\cite{Buras:1998raa, Buchalla_1996}. 
Typically this cancellation 
is precise through the order of the 
perturbative calculation. 
In this way, the operator product expansion provides a clean separation between short-distance perturbative physics and long-distance nonperturbative effects.

If the scale $\mu$ is chosen sufficiently high that the QCD coupling remains in the perturbative regime, the Wilson coefficients can be computed reliably using perturbative renormalization-group techniques. In contrast, the evaluation of the hadronic matrix elements $\langle \Theta_i(\mu) \rangle$, which probe long-distance physics, necessarily requires nonperturbative methods, such as lattice QCD (LQCD), chiral 
EFT~\cite{Gasser:1983yg,Gasser:1984gg}, or large-$N_c$ expansions~\cite{tHooft:2002ufq}.

\subsection{Effective weak Hamiltonian for HPV at a scale of 2 GeV~\label{sec:Hweak2}}

The goal of studying parity violation at low energies requires a careful treatment of the interplay between weak and strong interaction dynamics. The central challenge lies in computing the 
two-nucleon matrix elements relevant to hadronic and nuclear observables. One of the primary objectives of the LQCD 
program is to determine these matrix elements directly. However, current lattice calculations have not yet reached the precision necessary for quantitative comparison with experimental measurements. In parallel, chiral EFT 
approaches  provide a systematic low-energy framework, 
but these rely on experimental inputs for  effective couplings.
Consequently, state-of-the-art experimental results are often interpreted within the DDH framework of hadronic parity violation  ~\cite{DDH1980},
wherein NN 
interactions are framed in terms of one-meson exchanges. While pioneering, the DDH framework approximates QCD renormalization effects through phenomenological enhancement factors that neglects operator mixing~\cite{DDH1980}.

A natural \textit{ab initio} improvement in this regard is to construct the low-energy effective Hamiltonian governing flavor-conserving, parity-violating quark processes within the SM~\cite{Dai:1991bx,Kaplan:1992vj, Gardner_plb:2022mxf, Gardner_prc:2022dwi, Muralidhara:2023zcs}. To achieve this, the low-energy effects are obtained by evolving the effective Hamiltonian 
from high-energy scales down to hadronic scales using RG methods.
The resulting low-energy effective Hamiltonian can then be matched onto hadronic degrees of freedom to compute the parity-violating meson-nucleon couplings of DDH, 
thereby enabling a more direct, although still underconstrained, comparison with experimental extractions, given that the number of effective couplings exceeds the number of independent measurements. 
We note Secs.~\ref{sec:lqcd} and 
\ref{sec:chiral_few} for a discussion
of natural future steps in an 
{\it ab initio} analysis. 

As discussed in the previous section, the required effective Hamiltonian is an operator product expansion: 
\begin{equation}\label{Std HPV}
H^{\rm HPV}_{\rm eff}(\mu) = \frac{G_F \sin^2\theta_W}{3\sqrt{2}} 
\sum_{i=1} C_i (\mu) \Theta_i\ \,,
\end{equation}
where $C_i(\mu)$ are Wilson coefficients associated with $\Theta_i$ four-quark operators. 
Twelve such operators form a closed set under QCD mixing and describe 
the theory of hadronic parity violation ~\cite{Gardner_plb:2022mxf, Gardner_prc:2022dwi, Muralidhara:2023zcs}.
Moreover,  
$\theta_W$ is the weak-mixing angle, and the Fermi constant
$G_F\simeq 1.166 \times 10^{-5}\, \rm{GeV^{-2}}$~\cite{ParticleDataGroup:2024cfk}. At the scale of the $W$-boson mass, $M_W$, strong-interaction effects are perturbatively small. The parity-violating weak interaction is therefore obtained by matching the full SM
theory onto an effective theory in which the heavy electroweak gauge bosons
are removed as explicit dynamical degrees of freedom.
Summing over all flavor-conserving tree-level diagrams and isolating the parity-violating components of the amplitudes yields the set of Wilson coefficients $\vec{C}(M_W)$, 
and the 
parity-violating effective Hamiltonian 
at scale $M_W$.\footnote{With an overall factor of $\sin^2\theta_W$ extracted in Eq.\eqref{Std HPV}, the Wilson coefficients $\vec{C}(M_W)$ exhibit distinct dependence on the weak mixing angle: some contain a compensating $1/\sin^2\theta_W$ factor, while others do not. These distinguishing 
features dissolve into the 
different operators
due to RG-flow-induced operator-mixing in lower energy Wilson 
coefficients~\cite{Dai:1991bx, Gardner_plb:2022mxf}.}
To connect this short-distance description to hadronic scales, the effective Hamiltonian is evolved from the matching scale $M_W$ down to 
the scale 
$\mu \sim 2~\text{GeV}$ using RG 
methods. 
At next-to-leading order (NLO) in QCD, that is, including two-loop gluonic contributions, the Wilson coefficients evolve as~\cite{Muralidhara:2023zcs} (a similar structure holds for the LO evolution as well~\cite{Gardner_plb:2022mxf}):
\begin{equation}
    \Vec{C}(2 \, {\rm GeV}) =  U_4 (2\, {\rm GeV}, \mu_b) M(\mu_b,5) U_5 (\mu_b, M_W) \Vec{C}(M_W) \,.
\end{equation}
At the matching scale $\mu = M_W$,
the effective theory contains five active dynamical quark flavors. 
As the renormalization scale is lowered, heavy quark flavors are sequentially removed from the effective theory at their respective mass thresholds. In this way, the theory transitions from five to four active flavors at $\mu \sim m_b$, and from four to three active flavors at $\mu \sim m_c$. 
Continuity across these thresholds is ensured by imposing matching conditions:
with the input value at $Z^0$ mass 
scale, $\alpha_s(M_z) = 0.117$ \cite{ParticleDataGroup:2024cfk}, 
the resulting strong interaction 
strength ratios at NLO
are,
\begin{equation}
\begin{split}
  &\frac{\alpha_s(m_b=4.18\,\text{GeV} )}{\alpha_s(M_W=80.379\,\text{GeV} )}= 1.86;\, \quad \frac{\alpha_s(m_c=1.27\,\text{GeV} )}{\alpha_s(m_b)}= 1.75;\, \quad
  \frac{\alpha_s(\Lambda= 2\,\text{GeV} )}{\alpha_s(m_c)}=0.75\,.
  \end{split}
\end{equation}
The matching matrix $M(\mu,f)$ relates the Wilson coefficients in the $f$-flavor theory to those in the $(f-1)$-flavor theory at the threshold scale $\mu$. 
Between thresholds, the scale dependence of the Wilson coefficients is governed by the renormalization-group evolution matrix $U_f(\mu_1,\mu_2)$ within a fixed-$f$ effective theory,
and depends on the anomalous-dimension matrices $(\gamma(\alpha_s))$ computed at leading order and next-to-leading order in QCD:
\begin{equation}
    \gamma(\alpha_s) = \gamma_{\rm LO} + \gamma_{\rm NLO} \equiv \frac{\alpha _s}{4 \pi } \gamma^{(0)} + \left(\frac{\alpha _s}{4 \pi } \right)^2 \gamma^{(1)} \,.
\end{equation}
The anomalous dimension matrices ${\gamma}^{(0)}$ calculated in ~\cite{Gardner_plb:2022mxf} and ${\gamma}^{(1)}$ calculated in ~\cite{Muralidhara:2023zcs} encode, respectively, the leading-order  and next-to-leading-order  mixing and renormalization of the operator basis ${\Theta_i}$ under QCD. We refer
to \cite{Gardner_plb:2022mxf,Muralidhara:2023zcs,Muralidhara:2024thesis} for all details.

In practice, the perturbative renormalization-group evolution is typically carried down to a scale of order $\mu \sim 1 \, \mathrm{GeV}$~\cite{Dai:1991bx}. Below this scale, the strong coupling constant $\alpha_s(\mu)$ becomes numerically large, approaching values of order unity. Asymptotic freedom implies that QCD is weakly coupled at high energies, but becomes increasingly strongly coupled at low energies. Once $\alpha_s(\mu) \sim \mathcal{O}(1)$, the perturbative expansion in powers of $\alpha_s$ ceases to be reliable, and nonperturbative dynamics dominate.\\

\noindent\underline{Application to meson-nucleon parity-violating couplings}\\

Within the phenomenological framework of DDH,
the parity-violating meson ($M$)-nucleon ($N$) coupling constants of isospin $I$, conventionally denoted $h_M^I$,
are introduced to quantify the observable effects of hadronic parity violation.
In this model, these couplings enter 
via the hadron-level Hamiltonian, $\mathcal{H}_{\rm DDH}$.
The quark-level effective Hamiltonian under discussion here can be used to compute the same couplings in a robust way.

\begin{figure}[htp]
    \centering
    \includegraphics[width=8cm]{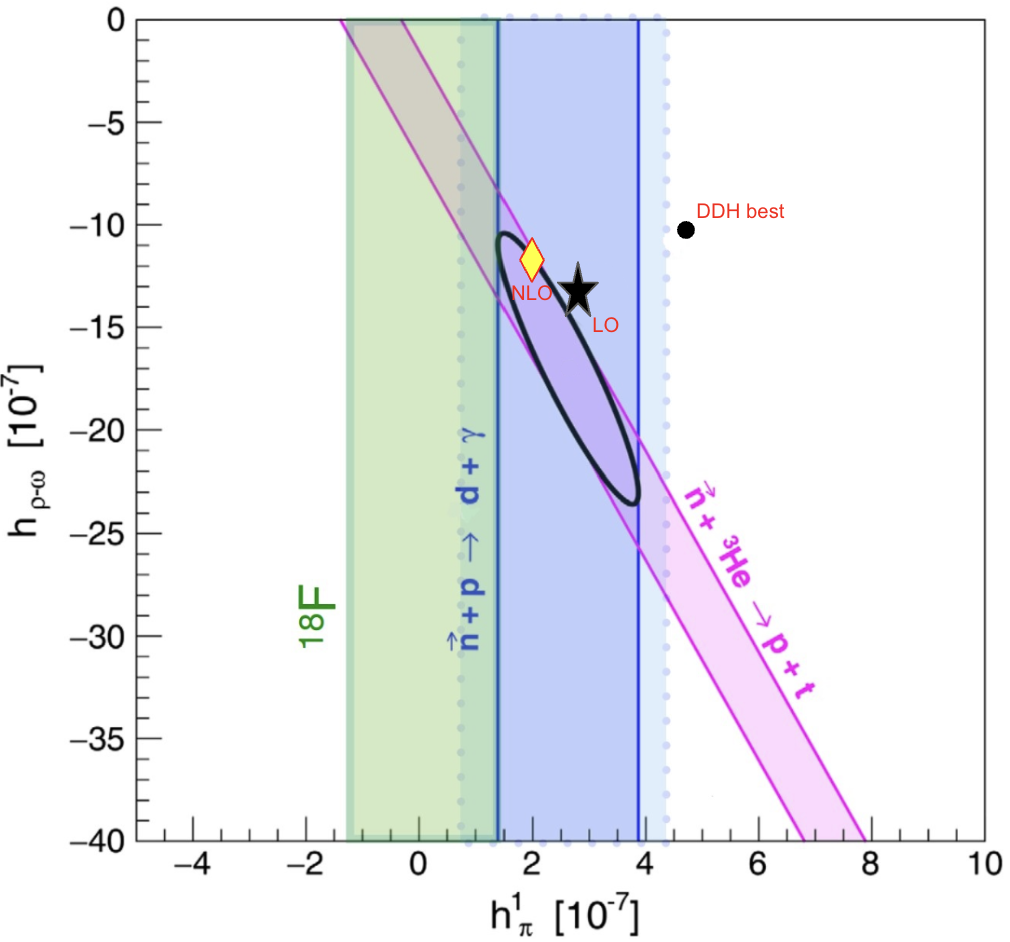}
    \caption{Different constraints on the 
    coupling constants $h_{\rho-\omega}$ and $h_{\pi}^1$ are compiled and compared.
    The vertical band bounded by a solid line
    is the
    value $h_{\pi}^1=(2.6 \pm 1.2) \times 10^{-7}$, 
    as measured in the parity-violating asymmetry in ${\vec n} + p\to d + \gamma$~\cite{NPDGamma:2018vhh}. 
    Its determination in 
    chiral perturbation theory,
    $h_{\pi}^1=(2.7 \pm 1.8) \times 10^{-7}$,
    is shown
    as the vertical band bounded by a dotted line~\cite{deVries:2015pza,deVries:2020iea}. 
    The diagonal constraint arises from the measured parity-violating asymmetry in $\vec{n} + ^{3}\mathrm{He} \to p + t$~\cite{Gericke2020}. A combined fit to the two experiments produces the ellipse shown. The analysis of $^{18}\mathrm{F}$ radiative decay provides the bound $|h_\pi^1| < 1.3 \times 10^{-7}$~\cite{Haxton:2013aca}, indicated by the leftmost vertical band. The phenomenological DDH \textit{best value}~\cite{DDH1980} is also included. The \textit{ab initio} results presented here, as described in text, at a scale of 2 GeV are denoted by a \textit{star} (LO) and a \textit{diamond} (NLO), with their sizes approximately reflecting the uncertainties from input parameters. The figure is reproduced from \cite{Muralidhara:2024thesis} with permission.
}
\label{NLO graph}
\end{figure}

Before proceeding to do so, it is insightful to separate the quark-level effective Hamiltonian into its isospin components
~\cite{Gardner_prc:2022dwi}: the odd sector with $I = 1$, and the even sector with $I = 0 \oplus 2$. This decomposition makes explicit the isospin structure of the underlying weak interaction,
\begin{equation}
    \mathcal{H}_{\rm eff}^{\rm PV}(2 \,\rm GeV) = \mathcal{H}_{\rm eff}^{\rm I=1}(2 \,\rm GeV) + \mathcal{H}_{\rm eff}^{\rm I=0\oplus2}(2 \,\rm GeV),
\end{equation}
where $\mathcal{H}_{\rm eff}^{\rm I=1}\,\, \rm{and}\,\, \mathcal{H}_{\rm eff}^{\rm I=0\oplus2}$ pack the contributions from both $W^{\pm}$ and $Z^0$ exchanges and themselves obey  the operator product expansion,
\begin{equation}
        \mathcal{H}_{\rm eff}^{I}(\mu) = \frac{G_F\sin^2\theta_W}{3\sqrt{2}}\sum_{i=1} C_i(\mu)^I\Theta_i^I(\mu) \,.
\end{equation}
Now, by matching the quark-level and hadron-level matrix elements: 
$\langle M N' | \mathcal{H}_{\rm eff}^I  (C^I_i) | N\rangle = \langle M N' | \mathcal{H}_{\rm DDH} (h_M^I)  | N\rangle$, these couplings can be estimated. As an example, phenomenologically, the pion contribution to hadronic parity violation
enters through the coupling $h_{\pi}^1$ with
    $\mathcal{H}^{\pi}_{\rm DDH} = i h^1_{\pi}(\pi^{+}\bar{p}n-\pi^{-}\bar{n}p)$.
Matching the quark and hadron-level matrix elements we have 
\begin{equation}
    -i h_{\pi}^1 \bar{u}_n u_p = \bra{n\pi^{+}}\mathcal{H}_{\rm eff}^{I=1}(C^1_i)\ket{p}\,,
    \label{pimatch}
\end{equation}
where $u_N$ with $N\in p,n$ is a Dirac spinor.
For the lack of
exact or 
fully non-perturbative evaluations of
hadronic matrix elements
$\braket{\Theta_i(\mu)}$,
the factorization approximation method~\cite{Fleischer:2010ca}
is used 
to split 
$\bra{n\pi^{+}}\mathcal{H}_{\rm eff}^{I=1}\ket{p}$
into separate pion-production and 
nucleon interaction terms.
Moreover, for the nucleon interaction evaluation, 
$\bra{n}\bar{d}u\ket{p}\equiv g_s^{u-d}\bar{u}_nu_p$,
precise LQCD
determinations of the 
scalar charge $g_s^{u-d}$
~\cite{Gupta:2018qil,Aoki:2021kgd}
are
employed. Finally at NLO precision, this yields~\cite{Muralidhara:2023zcs, Muralidhara:2024thesis}
\footnote{
Here we report the couplings 
accounting for the 
scheme dependent (`t Hooft-Veltman) corrections to the input NLO Wilson coefficients from LO diagrams,  coming from matching
the effective theory onto the full theory,
 as it appears in the appendix 
 \cite{Muralidhara:2024thesis}.}, 
\begin{equation}\label{hpihere}
h_{\pi}^{1} = (2.13 \pm 0.22) \times 10^{-7} \,\quad
    [\rm{LO}:\, (3.06 \pm 0.34) \times 10^{-7}~\cite{Gardner_plb:2022mxf,Gardner_prc:2022dwi}],
\end{equation}
with the 
leading-order results given in square brackets for comparison. The error estimates come from the LQCD inputs. The same coupling 
from the NPDGamma experimental determination is
$h_\pi^1 = (2.6 \pm 1.2_{\rm stat} \pm 0.2_{\rm sys})\times 10^{-7}$~\cite{NPDGamma:2018vhh}. The rest of the meson-nucleon couplings can be computed similarly. 
Their evaluations at NLO are:
\footnote{We note \cite{Phillips:2014kna,Schindler:2015nga} for 
assessments of the scaling of these couplings with $\sin^2\theta_W$ and with (large) $N_c$ to estimate their relative magnitude, but reserve comment 
to Sec.~\ref{sec:chiral_few}, as they come
from comparing the DDH and pionless EFT NN potentials.}
\begin{equation}\label{updatedDDH}
\begin{split}
 &h^1_{\rho} = (-0.260\pm 0.038) \times 10^{-7};\\
 &h^1_{\omega}= (1.51\pm 0.09 )\times 10^{-7};\quad
 h^0_{\omega} = (0.274\pm 0.013) \times 10^{-7}\\
 &h_{\rho}^{0} = (- 10.5\pm 0.6 )\times 10^{-7};\quad
 h_{\rho}^{2} = (9.59\pm 0.64)  \times 10^{-7}
\end{split}
\end{equation}
We can compare these determinations with 
an experimental constraint from 
the measurement of 
the parity-violating asymmetry 
in $\vec{n}+ {}^3{\rm He} \to t + p$. 
Specifically, the $n ^3$He collaboration
determines the 
combination of vector-meson-nucleon couplings,
$h_{\rho-\omega} \equiv h_\rho^0 + 0.605 h_\omega^0 - 
0.605 h_\rho^1 -1.316 h_\omega^1 + 0.026 h_\rho^2$ 
to be 
$h_{\rho-\omega}=(-17.0 \pm 6.56)\times 10^{-7}$~\cite{Gericke2020}. Using
the theoretical evaluations in 
Eq.~(\ref{updatedDDH}) yields 
$h_{\rho-\omega} = (-11.9\pm 0.65) \times 10^{-7}$ in NLO. We observe that both NLO results for $h_\pi^1$ and
$ h_{\rho-\omega}$ are within $\pm 1\sigma$ of the experimental determinations~\cite{Muralidhara:2024thesis}, 
as illustrated in Fig.~\ref{NLO graph}. 

Finally, it is important to note that the Wilson coefficients and the local operators depend on both the renormalization scale $\mu$ and the renormalization scheme adopted in computing the NLO QCD corrections. 
However, physical amplitudes computed using them must be independent of these purely calculational artifacts. 
In particular, the scheme and scale dependence of the Wilson coefficients $C_i(\mu)$ must cancel that of the corresponding hadronic matrix elements $\braket{\Theta_i(\mu)}$, up to higher-order corrections in perturbation theory~\cite{Buchalla_1996}. 
However, fully non-perturbative evaluations of $\braket{\Theta_i(\mu)}$ are not yet available.
As a consequence, the approximations and determinations described above retain a residual dependence on the renormalization scale and scheme.
But amplitudes such as the ones presented above at NLO values tend to be smaller than LO results and 
NLO results show less variation about the $\mu=2\,\rm GeV$ scale,
indicating 
the reduction of scale sensitivity in moving 
from LO to NLO.
For instance, LO and NLO estimates for $h_{\pi}^1$ at $\mu=2\,\mathrm{GeV}$ with variation between $\sim1\, \mathrm{GeV}$ 
(upper entry) to 
$4\, \mathrm{GeV}$ (lower entry),
\begin{equation}
\begin{split}
    \mathrm{NLO}: \, h_{\pi}^{1} &= 2.13 \pm 0.22 + \left(\stackrel{{+0.19}}{{}_{-0.33}}\right) \times 10^{-7} \quad 
    \mathrm{LO}:\,  3.06 \pm 0.34  + \left(\stackrel{{+1.29}}{{}_{-0.64}}\right)\times 10^{-7}.
\end{split} 
\label{hpi_scale}
\end{equation}

Before concluding this section, 
we pause to consider the broader outcomes of 
Fig.~\ref{NLO graph}. The LO and NLO
{\it ab initio} assessments of 
$h_\pi^1$ and $h_{\rho-\omega}$
are not compatible with the 
``DDH best'' assessment in 
\cite{DDH1980}, nor
should they be --- the possibility
of operator mixing, e.g., was not included
in that early analysis. 
The determined $h_\pi^1$ also appears to be 
in tension with the
upper bound reported in the 
analysis of $^{18}{\rm F}$~\cite{Haxton:2013aca}. 
We emphasize, however, that this
apparent tension may not be 
significant --- 
the physical scale of the 
nuclear experiment should be
much less than $2\, \rm GeV$.~\footnote{We note \cite{Schindler:2018irz} for 
an explicit illustration of ``fast'' RG running 
at low $\mu$ in pionless EFT.} 
We compare the outcomes of 
the DDH model with chiral 
effective theory in 
Sec.~\ref{sec:chiral_few}; here we can see that the existing 
$h_\pi^1$ assessments
are compatible. 
Finally, with 
the updated couplings of Eqs.~(\ref{hpihere},\ref{updatedDDH}) in hand, 
we return to the 
$^{133}$Cs anapole tension noted in Sec.~\ref{sec:intro}.
Comparisons of hadronic parity 
violating observables in many-body nuclei are 
often made in a loose version of a principal component analysis~\cite{Greenacre:2022PCA}, 
with each observable 
plotted as a function of two nominally 
dominant couplings, while varying the
subleading ones, typically over the 
``reasonable range'' of couplings given in \cite{DDH1980}, to yield bands in a plot 
of the two possible dominant couplings, as shown in 
\cite{Carlson2002,Haxton2002,Safronova2018}. 
The assumed dominant couplings, with their updated 
assessments at $\mu = 2 \, \rm GeV$ are 
\begin{equation}\label{eq:dom-coupl}
\begin{split}
    h_\pi^1 - 0.12 h_\rho^1 - 0.18 h_\omega^1 = (1.89 \pm 0.22)\times10^{-7} \,; \\
    -(h_\rho^0 + 0.7 h_\omega^0) = (10.4 \pm 0.61)\times10^{-7}  \,, 
    \end{split} 
\end{equation}
which appears more compatible with 
the $^{18}$F and $^{19}$F results 
than the DDH best value --- 
but they remain incompatible with the outcomes shown for the 
$pp$ 
scattering
and $^{133}$Cs anapole determinations~\cite{Haxton:2013aca,Safronova2018}. 
The various results do rely on the values of the assumed sub-leading interactions, 
noting that the effect of two-pion exchange (TPE) has not been included here~\cite{Viviani2014}, 
and in the anapole case assessments of nuclear many-body correlations and of atomic effects are needed to make a comparison at all. 
We discuss this further 
in Sec.~\ref{sec:apv-anapole}, within this broader context.

\subsection{Parity violation in LQCD
\label{sec:lqcd}} 

LQCD 
is an {\em ab initio} method  
that allows for calculations at the scale pertinent to a problem under consideration. 
In LQCD, one evaluates matrix elements in Euclidean time using stochastic
methods in a finite volume $V$, at finite lattice spacing $a$ and at given values for the light quark masses $m_q$, which are often taken 
to be heavier
than the physical ones. Suitably tailored effective field theories are used to approach the limits $V\to\infty$, $a\to 0$ and $m_q\to m_q^{\rm phys}$, where $m_q^{\rm phys}$ denotes the physical value of the quark masses.
To assess the systematic uncertainties underlying 
LQCD, 
it is now common
practice to consider a certain number of volumes and lattice spacings,
see, e.g., the discussion in~\cite{FlavourLatticeAveragingGroupFLAG:2024oxs}.

So far, the parity-violating pion-nucleon coupling $h_\pi^1$ has been
considered in this approach. Conventionally, one works out the corresponding
three-point function, here the coupling of the weak current to the nucleon
at zero momentum transfer. Such a calculation was pioneered in~\cite{Wasem:2011tp}. This calculation was performed at one volume,
one lattice spacing and one unphysical pion mass. Also, it did not
include non-perturbative renormalization of the bare parity-violating operators, a chiral extrapolation to the physical pion mass, or contributions from disconnected (quark-loop) diagrams, so the resulting
$h_\pi^{1, \rm con} =(1.099 \pm 0.505^{+0.058}_{-0.064})\times 10^{-7}$
should only be considered as an approximate result.

A different approach to the problem was laid out in~\cite{Feng:2017iqb}.
There, current algebra (or leading order chiral perturbation theory) was used to map the parity-violating pion emission of a nucleon on a parity-conserving (PC) single nucleon matrix element, $\lim_{p_\pi\to 0}\langle n\pi^+ | {\cal L}^w_{\rm PV}(0)|p\rangle \simeq -\sqrt{2} i\langle p | {\cal L}^w_{\rm PC} (0) | p\rangle/F_\pi$, where corrections to this leading order
relation are discussed in~\cite{Guo:2018aiq}. The parity-conserving matrix element can be mapped on a four-quark mass shift of the nucleon mass,
$(\delta m_N)_{4q}= \langle p | {\cal L}^w_{\rm PC} (0) | p\rangle/m_N$,
so that
\begin{equation}
h_\pi^1 \simeq -\frac{1}{F_\pi}\frac{(\delta m_N)_{4q}}{\sqrt{2}}~.  
\end{equation}
It can be shown that  $(\delta m_N)_{4q}$ is nothing but the neutron-proton mass splitting induced by ${\cal L}^w_{\rm PC}$.
This alternative theoretical ansatz leads to a major simplification in the lattice computation as one now considers a transition amplitude between single nucleon states with a parity-conserving Lagrangian. From a numerical point of view this is more straightforward to handle in a lattice calculation. In particular, the complication arising from the pion-nucleon state is absent since the matrix element is computed for single
nucleon initial and final states. This task has been taken up by the Bonn LQCD group in~\cite{Petschlies:2023kly}, where  a concrete
numerical implementation to evaluate the nucleon
3-point functions with the 4-quark operator insertions
of ${\cal L}^w_{\rm PC}$ was proposed and carried out (again at one lattice spacing, one volume, and one value for the light quark mass). 
The issue of renormalization is still under construction, so only a bare value of $h_\pi^1$ was reported, $h_\pi^{1, {\rm bare}} = 8.08 (98) \times 10^{-7}$.
This is on the large side of the phenomenological values discussed above, but 
we note again that this value will change after renormalization. This poses a formidable challenge due to the mixing with lower dimensional operators in the twisted mass formulation employed in~\cite{Petschlies:2023kly} (see also that paper for much more detail). It will thus be very interesting to see
what the full calculation that is under way will give.

Before concluding this section, 
we pause to consider 
the possibility of calculating 
parity-violating NN matrix elements, 
as this could yield key complementary information 
to the $h_\pi^1$ studies we have noted. 
For example, it has long been thought that the 
parity-violating 
$|\Delta I|=2$ NN matrix element, due to its 
lack of an annihilation channel, 
would be the most suitable first step
in such a campaign~\cite{Kurth_LQCD-NN-isotensor:2015cvl}, but complexities 
in identifying whether a NN system in LQCD 
is bound or not require resolution 
first~\cite{Walker-Loud22}. 
Great progress on this latter problem has been 
made recently~\cite{BaryonScattering:2025ziz,Moscoso-NN_LQCD:2026wmz}, and we hope 
parity-violating NN studies in LQCD can resume. 

\subsection{Parity violation in chiral effective theory (few nucleons)
\label{sec:chiral_few}}

Chiral nuclear EFT (ChEFT) is the method of choice to investigate parity violation in few-nucleon systems.
For a detailed introduction, see~\cite{Meissner:2022cbi}.
This approach has a number of advantages over the one-boson-exchange model discussed before. First,
it directly reflects the (broken) symmetries of the SM. 
Second,  the symmetry-conserving and -violating interactions among the relevant degrees of freedom, pions and nucleons, are obtained within the same framework which allows for consistent calculations. Such calculations can be improved order by order in a controlled expansion within the chiral EFT. Here, we discuss the parity-violating (PV) potentials up to next-to-next-to-leading order (N$^2$LO) as they have a richer structures than the  one-boson-exchange (OBE) model \cite{DDH1980}. Third, the chiral EFT approach can be extended to multi-nucleon or electromagnetic interactions allowing for a unified treatment of various different observables. The main
disadvantage is the appearance of a number of low-energy constants (LECs). These can either be determined from fits to data
(if the data base 
is large enough) or needs to be calculated on the lattice or using the methods described above. 
For reviews on ChEFT applied to hadronic parity-violations, see 
\cite{Holstein:2009zzb,deVries:2015gea,deVries:2020iea}.
Note that another EFT exists, that can be used to describe
hadronic parity violation and other low energy observables. This is the so-called
pionless EFT. In that approach the pion is integrated out and all nuclear interactions are described by NN, 3N ,... contact terms. We do not consider this in detail here, but refer the reader to the review~\cite{Schindler2013} and the recent work in~\cite{Nguyen:2020quk}.\\ 

\noindent \underline{Parity-Violating Two-Nucleon Potential}\\

 At a scale slightly below the mass of the $W$ boson, the PV operators involving the light $u$ and $d$ quarks can be  written as
\begin{equation}\label{FQPV}
\mathcal L_{\mathrm{PVTC}} = \frac{G_F}{\sqrt{2}} \bigg[ \left(\frac{1}{2}-\frac{1}{3} \sin^2 \theta_W\right)\,V_\mu^a A^{\mu a}  - \frac{1}{3}\sin^2 \theta_W\,I_\mu A^{\mu3}
-\sin^2 \theta_W \left(V_\mu^3 A^{\mu 3}-\frac{1}{3}V_\mu^a A^{\mu a} \right)\bigg]+\dots\,\,\,,
\end{equation}
in terms of the currents $V_\mu^a = \bar q \gamma_\mu \tau^a q$,
$A_\mu^a = \bar q \gamma_\mu \gamma^5 \tau^a q$, and $I_\mu = \bar q
\gamma_\mu q$, where we use the the quark doublet $q = (u\,d)^T$ and
$\tau^a$ is the weak isospin operator. Note that we have set the
Cabibbo angle to unity. The dots denote operators involving heavier
quarks. In particular, operators with strange quarks can have
important consequences for nuclear PV effects. The Lagrangian in
Eq.~\eqref{FQPV} needs to be brought to lower energies via
RG evolution which dresses the coupling constants with $\mathcal O(1)$ QCD factors and induces operators with different color structure \cite{Dai:1991bx,Gardner_plb:2022mxf,Tiburzi:2012hx,Muralidhara:2023zcs}. (We note 
\cite{Muralidhara:2023zcs} supercedes 
\cite{Tiburzi:2012hx}.) The ChEFT Lagrangian is then obtained by constructing all chiral-invariant interactions and all interactions that break chiral symmetry in the same way as the chiral-breaking sources at the quark level. This leads to an infinite tower of interactions, but these are ordered by the chiral index $\Delta = d + n/2-2$, where $d$ counts the number of derivatives or pion mass insertions and $n$ the number of fermion (nucleon) fields at a given vertex. 
Any derivative amounts to a small momentum $q$, and 
the pion mass $M_\pi$ is also small. Consequently, $q/\Lambda_\chi$ and $M_\pi/\Lambda_\chi$, with $\Lambda_\chi \simeq 1\,$GeV the chiral symmetry breaking
scale, are the small parameters underlying the chiral power counting of our EFT. The nucleon mass $m_N$ that is of the
same size as $\Lambda_\chi$ requires a special treatment as detailed in~\cite{Meissner:2022cbi}.
The three four-quark interactions transform, respectively, as a chiral singlet, an isovector ($\Delta I=1$), and an isotensor ($\Delta I=2$) \cite{Kaplan:1992vj}. This implies that the first operator can only induce pionic operators with derivatives, while the other two can lead to non-derivative pionic interactions. Only the isovector interaction leads to an interaction with chiral index $\Delta =-1$:
\begin{equation}
\mathcal L_{\mathrm{PV}}^{(-1)} = \frac{h_\pi^1}{\sqrt{2}}\Nb(\vec \pi \times \vec\tau)^3 N\,\,\,,
\end{equation}
in terms of the already discussed weak pion-nucleon coupling constant $h_\pi^1$. Two remarks are in order: First, only charged
pions are involved in this interaction, in agreement with Barton's theorem~\cite{Barton:1961eg}. Second, the chiral 
dimension is lower than for the PC case, where the OPE contributes at $\Delta=0$. This does not imply that the
PV interaction is larger than the PC one, as its effects are greatly suppressed by the prefactor $G_F$. The chiral index is simply used as a tool to organize the various terms in the expansion in small momenta and the pion mass.
Also, different to the PC case, there appear no $N\!N$ 
contact interactions at LO. Such terms require at least one derivative and have chiral index $\Delta =1$. Thus, the LO PV potential consists only of the OPE contribution:
\begin{equation}
V^{(-1)}_{\text{PV}}
= - \frac{g_{A}h_\pi^1}{ 2\sqrt{2} F_\pi} i(\vec \tau_1\times \vec \tau_2)^3 \frac{(\vec \sigma_1+\vec \sigma_2)\cdot \vec q }{\mpi^2+q^2}\,\,\,,
\label{onepion}
\end{equation}
in terms of the nucleon spin $\vec \sigma_{1,2}$ and the momentum transfer flowing from nucleon $(1)$ to nucleon $(2)$: $\vec q = \vec p - \vec p^{\,\prime}$ ($q = |\vec q\,|$), where $\vec p$ and $\vec
p^{\,\prime}$ are the relative momenta of the incoming and
outgoing nucleon pair in the center-of-mass frame. Further, $F_\pi = 92.1\,$MeV is the weak pion decay constant, $g_A =1.275$ the nucleon's axial-vector coupling and $M_\pi =139.57\,$MeV the charged pion mass~\cite{ParticleDataGroup:2024cfk}. Because this LO potential consists of a single term, one might naively expect that this term dominates hadronic and nuclear PV. From the existing data it should then be possible to fix the size of $h_\pi^1$ from which other processes can be predicted. Unfortunately, the situation is more complicated for two main reasons: First, the LO potential changes the total isospin of the interacting nucleon pair and therefore does not contribute to PV effects in proton-proton ($pp$) scattering. As a significant part of the nonzero PV measurements has been made in this process, higher-order corrections are required to analyze the data. Second, the division of the potential into LO, next-to-leading order (NLO), .... , is based on an expansion in $p/\Lambda_\chi$. However, the isovector four-quark operator is suppressed by a factor $\sin^2 \theta_W \sim 1/5$. Large $N_c$ arguments indicate that $h_\pi^1$ could even be further suppressed \cite{Kaiser:1989ah,Meissner:1998pu, Phillips:2014kna}. Thus, formally higher-order corrections might be larger than expected because of dimensionless factors not captured by the chiral counting.

A great advantage of ChEFT is that higher-order corrections can be
systematically calculated. The NLO potential was first obtained
in~\cite{Zhu2005}
and shown to consist of TPE 
diagrams proportional to $h_\pi^1$ and $N\!N$ contact interactions \cite{Girlanda:2008ts}. 
The TPE contributions suffer from UV divergences that are absorbed, together with the associated scale dependence, by the contact terms. One way to handle these divergences in the PV potential is 
through the so-called spectral function regularization with a cut-off $\Lambda_S$, see~\cite{Epelbaum:2003gr} for the PC potential. In this way, the PV and PC potentials are regularized in the same way. By taking the spectral cut-off $\Lambda_S\rightarrow \infty$, the results in dimensional regularization are retrieved. The TPE contributions are then given by~\cite{Zhu2005,Kaiser:2007zzb,deVries:2014vqa}
\begin{eqnarray}\label{NLOTPE}
V^{(1)}_{\text{PVTC}}
(\Lambda_S)
&=&   -\frac{ g_A h_\pi^1}{2 \sqrt{2}\Fp} \frac{1}{(4\pi \Fp)^2}\left[i (\vec \tau_1\times \vec \tau_2)^3 (\vec \sigma_1+\vec \sigma_2)\cdot \vec q\right] \left(g_A^2 \frac{8\mpi^2+3q^2}{\omega^2} - 1 \right) L(q,\Lambda_S)\nonumber\\
&& +  \frac{ g_A^3 h_\pi^1}{2 \sqrt{2}\Fp}  \frac{4}{(4\pi \Fp)^2}\left[i (\vec \tau_1+ \vec \tau_2)^3 (\vec \sigma_1\times \vec \sigma_2)\cdot \vec q\right] L(q, \Lambda_S)\,\,\,,
\end{eqnarray}
in terms of the loop functions
\begin{equation}
\w = \sqrt{q^2+4\mpi^2}\,\,, \hspace{3mm} L(q,\Lambda_S)= \frac{\w}{2q} \log\left(\frac{\Lambda_S^2 \w^2 +q^2 s^2 + 2 \Lambda_S s \w q}{4\mpi^2(\Lambda_S^2+q^2)}\right)\,\,,\hspace{3mm} s = \sqrt{\Lambda_S^2-4\mpi^2}\,\,\,.
\end{equation}
The first term in Eq.~\eqref{NLOTPE} has the same spin-isospin structure as the OPE potential and is therefore not very interesting. However, the second term induces ${}^1S_0\leftrightarrow {}^3 P_0$ transitions and gives the first contributions to $pp$ scattering. 

\begin{figure}[t!]
    \centering
    \includegraphics[width=0.5\linewidth]{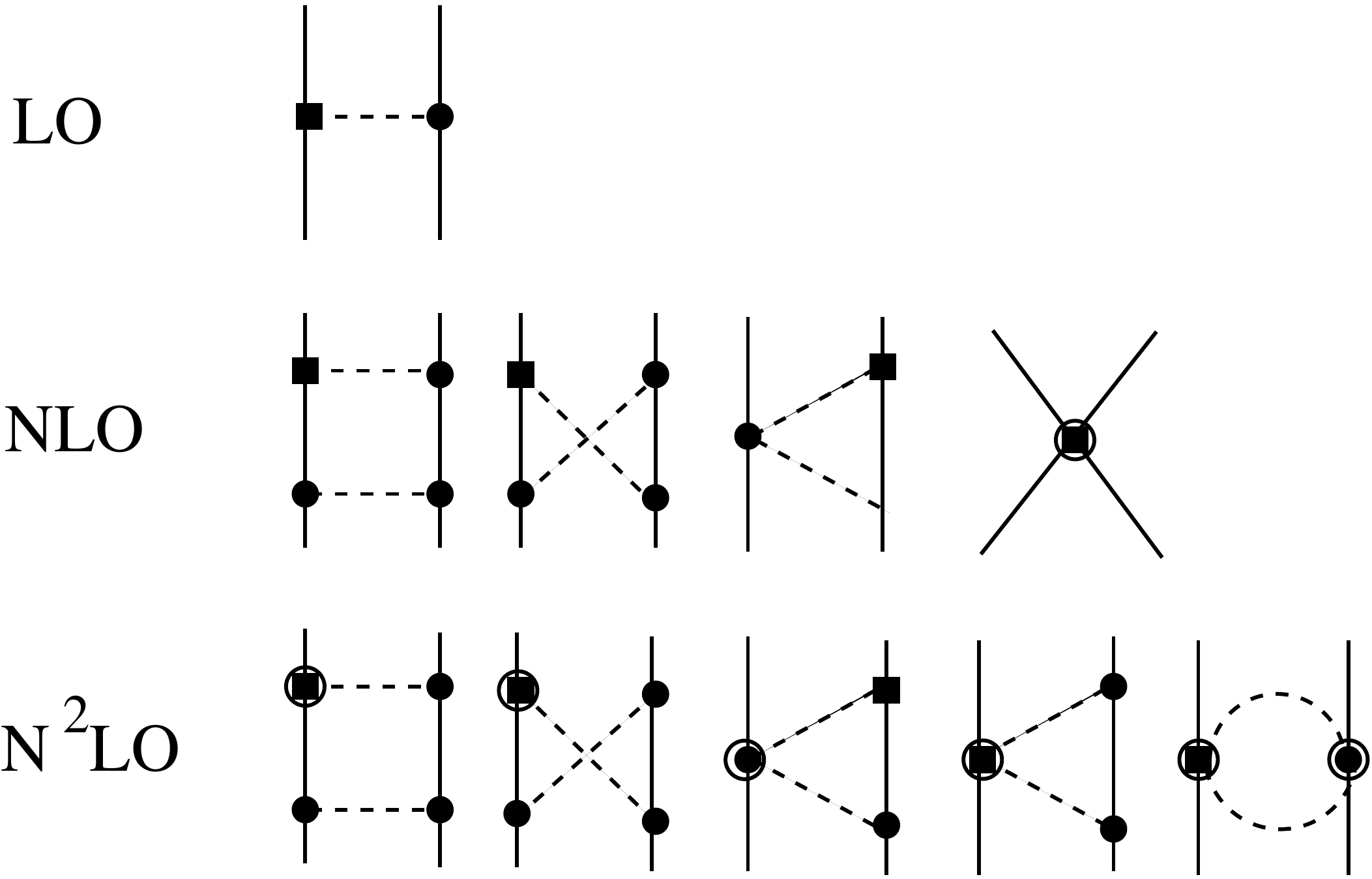}
    \caption{Chiral expansion of the  PV  NN 
    force up to N$^2$LO. Solid and dashed lines denote nucleons and pions, respectively. Circles and squares represent, respectively, LO PC and PV interactions while circled vertices represent NLO interactions. Corrections to the one-pion-exchange potential and three-body forces are not shown.}
    \label{FigPot1}
\end{figure}

At the same order as the TPE diagrams, one has five\footnote{This number can be understood by noticing that there are five possible $S\leftrightarrow P$ couplings. One ${}^3S_1 \leftrightarrow {}^1 P_1$ transition, one ${}^3S_1 \leftrightarrow {}^3 P_1$ transition, and three, one for each value of the change in isospin 
$\Delta I =0,1,2$ in 
${}^1S_0 \leftrightarrow {}^3 P_0$ transitions.}
 $N\!N$ contact interactions. These can be written in various ways \cite{Girlanda:2008ts, Phillips:2008hn}, and here we use the following parametrization \cite{deVries:2014vqa}
\begin{equation}\label{contactPV}
V^{(1)}_{\text{PVTC}}
=  \frac{C_0}{ \Fp \Lambda_\chi^2}  (\vec \sigma_1 - \vec \sigma_2)\cdot (\vec p + \vec p^{\,\prime} ) + \frac{1}{ \Fp \Lambda_\chi^2} \left(C_1 + C_2\frac{(\vec\tau_1+\vec\tau_2)^3}{2} + C_3\frac{\vec \tau_1\cdot \vec \tau_2-3 \tau_1^3 \tau_2^3 }{2} \right) i(\vec \sigma_1 \times \vec \sigma_2) \cdot \vec q
 + \frac{C_4}{ \Fp \Lambda_\chi^2} i(\vec \tau_1\times \vec \tau_2)^3(\vec \sigma_1 + \vec \sigma_2)\cdot \vec q\,\,\,.
\end{equation}
All together, the NLO PV potential depends on six LECs which need to be fitted to experiments or calculated with nonperturbative techniques. At next-to-next-to-leading order (N$^2$LO) several additional TPE diagrams appear, which have been calculated in~\cite{deVries:2014vqa}. The first part involves no new LECs and is proportional to $\pi h_\pi^1\,c_4$, where $c_4$ arises from the parity-conserving 
$\pi\pi N$ vertex~\cite{Bernard:1995dp}, and has the same spin-isospin properties as the second term in Eq.~\eqref{NLOTPE}. Because of the large size of $c_4 \simeq 3.6\,$GeV${}^{-1}$ \cite{Hoferichter:2015tha}, explained by underlying $\Delta$ and $\rho$-meson resonances~\cite{Bernard:1996gq}, and the enhancement by a factor of $\pi$, this term can be expected to dominate the N$^2$LO potential unless $h_\pi^1$ is very small. The second part depends on five new PV pion-nucleon and pion-pion-nucleon LECs \cite{Kaplan:1992vj}, which will be difficult to fit to the scarce data. At this order we encounter the first contributions to PV three-body forces which only been studied in pionless EFT, see~\cite{Griesshammer:2010nd}, depending on new LECs. The hierarchy of the PV potential is sketched in  Fig.~\ref{FigPot1}.\\

\noindent \underline{Estimation of the LECs and mapping on the DDH approach}\\

Although ChEFT allows the determination of the hierarchy and form of the potential, the LECs cannot be obtained from symmetry considerations alone. The simplest estimates for these couplings are obtained from naive-dimensional analysis (NDA)\cite{Manohar:1983md, Weinberg:1989dx}. NDA predicts $h_\pi^1 \sim C_i \sim \mathcal O(G_F F_\pi \Lambda_\chi) \sim 10^{-6}$.
This is nothing but an order-of-magnitude estimate which roughly probes the size of PV in nuclear systems.

The four-nucleon LECs $C_i$ listed in Eq.~\eqref{contactPV} must either be determined by a fit or using the
resonance saturation method for four-nucleon operators developed in~\cite{Epelbaum:2001fm}. In the OBE model,
hadronic PV is generated through due to the exchange of a single pion, $\rho$-, or $\omega$-mesons, with one PV and one PC vertex in each meson-exchange. Integrating
out the contributions from the heavy mesons ($\rho,\omega$) allows to represent the LECs of the contact terms in the
following way (using the DDH potential in the form given in~\cite{deVries:2014vqa})
\begin{eqnarray}\label{comparison}
\frac{C_0 + C_1}{ \Fp\Lambda_\chi^2}  &\sim&\frac{1}{m_N}\left[\frac{g_\omega h^0_\omega \kappa_S}{m_\omega^2}c_{\omega}(0,\Lambda_{\omega}) - \frac{3 g_\rho h^0_\rho\kappa_V}{m_\rho^2} c_{\rho}(0,\Lambda_{\rho})\right]\,\,\,,\nonumber\\
\frac{-C_0 + C_1}{ \Fp\Lambda_\chi^2} &\sim&\frac{1}{m_N}\left[ \frac{g_\omega h^0_\omega (2 +\kappa_S)}{m_\omega^2}c_{\omega}(0,\Lambda_{\omega}) + \frac{g_\rho h^0_\rho (2 +\kappa_V)}{m_\rho^2}c_{\rho}(0,\Lambda_{\rho}) \right]\,\,\,,\nonumber\\
\frac{C_2}{ \Fp\Lambda_\chi^2}  &\sim&\frac{1}{m_N}\left[ \frac{g_\omega h^1_\omega (2 +\kappa_S)}{m_\omega^2}c_{\omega}(0,\Lambda_{\omega}) + \frac{g_\rho h^1_\rho (2 +\kappa_V)}{m_\rho^2} c_{\rho}(0,\Lambda_{\rho})\right] -  \frac{ g_A^3 h_\pi^1}{2 \sqrt{2}\Fp}  \frac{8}{(4\pi \Fp)^2} \frac{s}{\Lambda_S}\,\,\,,\nonumber\\
\frac{C_3}{ \Fp\Lambda_\chi^2} &\sim&-\frac{1}{m_N}\frac{g_\rho h^2_\rho (2 +\kappa_V)}{\sqrt{6}\, m_\rho^2}c_{\rho}(0,\Lambda_{\rho})\,\,\,,\nonumber\\
\frac{C_4}{\Fp\Lambda_\chi^2} &\sim&\frac{1}{2m_N}\left[\frac{g_\omega h^1_\omega }{m_\omega^2}c_{\omega}(0,\Lambda_{\omega}) + \frac{ g_\rho( h^{1\,\prime}_\rho-h^1_\rho)}{m_\rho^2} c_{\rho}(0,\Lambda_{\rho})\right] +  \frac{ g_A h_\pi^1}{2 \sqrt{2}\Fp}  \frac{(2g_A^2-1)}{(4\pi \Fp)^2} \frac{s}{\Lambda_S}\,\,\,.
\end{eqnarray}
in terms of the PC vertices with couplings $g_\omega =8.4 $, $g_\rho = 2.8$, $\kappa_S=-0.12$, and $\kappa_V = 3.70$, six PV meson-nucleon couplings $h_\rho^{0,1,2}$, $h_\rho^{1\,\prime}$, and $h_\omega^{0,1}$. Following \cite{Holstein:1981cg,Haxton:2013aca,deVries:2014vqa} we set $h_\rho^{1\,\prime}=0$, although it might not
be so small~\cite{Phillips:2014kna,Schindler:2015nga}.  
Also, the meson-baryon
vertices require regularization, expressed here in terms of the one-meson-exchange functions
\begin{equation}\label{DDHcutoff}
c_{\rho,\omega}(q^2,\Lambda_{\rho,\omega}) = \left(\frac{\Lambda_{\rho,\omega}^2 - m_{
\rho,\omega}^2}{\Lambda^2_{\rho,\omega}+q^2} \right)^2\,\,\,,
\end{equation}
where $m_\rho \simeq m_\omega \simeq 780$~MeV are, respectively, the masses of the $\rho$- and $\omega$-meson. Note that
the last terms contributing to $C_2$ and $C_4$ stem from the TPE. Using the values from DDH or the soliton model or the
updated couplings given in Eqs.~(\ref{hpihere},\ref{updatedDDH}) using NLO pQCD and LQCD for the quark flavor charges, one finds that the $C_i$ are a few times $10^{-6}$, consistent with the  NDA estimates, see Table~\ref{tab:Ci}. For comparison,
we also give the $C_i$ based on the best DDH values, as well as their
reasonable ranges~\cite{DDH1980}, and from the topological soliton model. 
From the updated couplings we see a pattern
in the isospin structure of the $C_i$, with 
those of the $I=0$ sector ($C_0, C_1$) and 
the $I=2$ sector ($C_3$) being of opposite sign, 
yet all being grossly larger in magnitude than 
the $I=1$ couplings ($C_2, C_4$). 
The dominance of the $I=0,2$ $C_i$ couplings 
bears favorable comparison with the outcomes of the 
$\sin^2\theta_W$ and large $N_c$ scaling analysis 
of \cite{Phillips:2014kna,Schindler:2015nga} 
from the pionless EFT description of the 
parity-violating NN potential, though that analysis does not capture the noted
relative sign, nor the phenomenological 
importance of the parity-violating one pion 
exchange contribution. Also the  
$\sin^2 \theta_W$ suppression~\cite{Phillips:2014kna,Schindler:2015nga} in the $h_\rho^2$ relative to the $h_\rho^0$ coupling does not seem to appear. This 
could be a consequence of 
operator mixing in the $I=0 \oplus 2$ sector from the $M_W$  
to the $\mu=2$ mass scale, as analyzed in 
\cite{Gardner_plb:2022mxf,Muralidhara:2023zcs} 
and discussed in Sec.~\ref{sec:Hweak2}.
\begin{table}[htb]
\begin{center}
\begin{tabular}{|c|cccc|}
\hline
  LEC   & This work & DDH best & DDH range & Soliton model \\
  \hline
  $C_0$ & 3.94 $\pm$ 0.23  & 4.58    &  
 [$-$0.8, 13.1] &  0.9 \\
  $C_1$ & 1.31 $\pm$ 0.07   & 1.16    &  
  [0.7,2.6]
  & 0.1 \\
  $C_2$ & $-$0.58 $\pm$ 0.06   & $-$2.26 &  [$-$4.8,$-$0.6] 
  & $-$0.7\\
  $C_3$ & $-$1.01 $\pm$ 0.07   & 1.00    &  
  [0.8,1.2] 
  & 0.4 \\
  $C_4$ & 0.26 $\pm$ 0.01  & 0.26  &  
  [$-$0.1,0.7] 
  & $-$0.05 \\
  \hline
\end{tabular}
\end{center}
\caption{Comparison of the four-nucleon LECs $C_i$ (in units of $10^{-6}$) using the weak meson-nucleon couplings of this work, 
see Eqs.~(\ref{hpihere},\ref{updatedDDH}), the best values from DDH~\cite{DDH1980} and their ranges (as reported in \cite{deVries:2014vqa}
with $F_\pi$, $g_A$, $M_\pi$ as noted in text and 
$m_N=938.92\, \rm MeV$.) 
as well as the topological soliton model of the nucleon~\cite{Kaiser:1989ah,Meissner:1998pu}.
For ease of 
comparison, we have set $c_{\rho,\omega}=1$ here together with $\Lambda_S =0.6$~GeV and $\Lambda_\chi =1$~GeV. 
\label{tab:Ci}}
\end{table}

\noindent \underline{Application to Few-Nucleon Processes}\\

Arguably the purest process to investigate hadronic PV in few-nucleon scattering is proton-proton ($pp$) scattering.
The longitudinal asymmetry (LAP) in $pp$ scattering  is defined as the difference in cross section of scattering between an unpolarized target and a beam of positive and negative helicity, normalized to the sum of cross sections. The existing experiments measured the LAP over a certain angular range (from $\theta_1$ to $\theta_2$) and report the integrated asymmetry:
\begin{equation}
\bar A_L(E,\theta_1,\theta_2) = \frac{\int_{\theta_1}^{\theta_2} d\Omega(d\sigma_+ - d\sigma_-)}{\int_{\theta_1}^{\theta_2} d\Omega(d\sigma_+ + d\sigma_-)}\,\,\,.
\end{equation}
It has been measured at several energies. The experiments with highest precision are the Bonn experiment at 13.6~MeV \cite{Eversheim:1991tg}, the PSI experiment \cite{Kistryn:1987tq}
at 45~MeV, and the TRIUMF experiment at 221~MeV \cite{TRIUMFE497:2001hga} (all energies are lab energies). Note that the
LO PV potential causes a ${}^3S_1 \leftrightarrow {}^3 P_1$ transition that is forbidden for 
two identical protons. Thus, one has to go at least to NLO. At this order, one has to deal with two LECs, namely
$h_\pi^1$ and $C$, which is a linear combination of the $C_i$ discussed before, $C =-C_0+C_1+C_2-C_3$, and the
corresponding operator induces the required ${}^1S_0 \leftrightarrow {}^3P_0$ transition. Analyzing these data at NLO, one 
finds $h_\pi^1 = (1.1\pm 2 )\cdot 10^{-6}$ and $C = (-6.5\pm 8)\cdot 10^{-6}$~\cite{deVries:2013fxa}.  The uncertainty in these
LECs is sizeable as only three data points exist. At N$^2$LO, one further combination of LECs related to higher-order PV pion-nucleon couplings appears. Its quantitative role very much depends on the actual smallness of $h_\pi^1$. Assuming 
natural sized couplings, the contribution of these higher order corrections stays well in the bounds suggested by the
power counting, however, if $h_\pi^1$ is as small as $10^{-7}$, these corrections will become sizeable. For a more detailed
discussion see~\cite{deVries:2014vqa}.

In contrast to PV $pp$ scattering, radiative neutron  capture on the proton, $\vec{n}p\to d\gamma$, where $\vec{n}$ denotes
a longitudinally polarized neutron and $d$ the deuteron, is sensitive to the leading PV OPE potential. 
The longitudinal analyzing power for this process is defined as 
\begin{equation}
 A_\gamma(\theta)  = \frac{d\sigma_+(\theta) - d \sigma_-(\theta)}{d\sigma_+(\theta) + d \sigma_-(\theta)} = a_\gamma\,\cos \theta\,\,\,,   
\end{equation}
with $d\sigma_\pm(\theta)$ the differential cross section for neutrons with positive/negative helicity and $\theta$ the angle between the photon momentum and the neutron spin. 
An analysis within
ChEFT indeed gives for the longitudinal analyzing power \cite{deVries:2015pza}
\begin{equation}
a_\gamma = (-0.11 \pm 0.05) h_\pi^1 - (0.5 \pm 0.5) \times 10^{-8}~,   
\end{equation}
where the second term stems from the short-distance contribution $\sim C_4$. Ignoring this latter small contribution
and comparing to the experimental value from~\cite{NPDGamma:2018vhh} gives $h_\pi^1 = (2.7\pm 1.8) \times 10^{-7}$,
which is consistent with the values discussed in  Sec.~\ref{sec:Hweak2}.

A few more observables have been investigated in chiral EFT~\cite{Viviani2014}. 
These are the spin rotation in $\vec n p$ or $\vec n d$ and the longitudinal analyzing power in the reaction $\vec n + {}^3\mathrm{He}\rightarrow  p +{}^3\mathrm{H}$. These are given in terms of $h_\pi^1$ and the leading order contact terms as
\begin{eqnarray}
\frac{d\phi}{dz}(\vec np)&=& (1.31\pm0.05)h_\pi^1 + (0.20\pm 0.01) C_0 -(0.23\pm0.01) C_1 - (0.44\pm0.01)C_3 - (0.09\pm0.01)C_4\nonumber\\
\frac{d\phi}{dz}(\vec nd)&=& (2.20\pm0.02)h_\pi^1 - (0.08\pm 0.01) C_0 - (0.19\pm0.01)C_4\nonumber\\
A_L &=& -(0.14\pm0.01)h_\pi^1 + (0.017\pm 0.003) C_0 -(0.007\pm 0.001) C_1+ (0.008\pm0.001) C_2 + (0.018\pm0.002)C_4~.
\end{eqnarray}
 The $h_\pi^1$ dependence is dominated by the OPE potential, with TPE contributions entering at the $10\%-30\%$ level as expected from power counting. 
The spin rotation angles have not been measured yet, but new experiments are being considered
or are under development, as noted in 
Secs.~\ref{sec:expt},\ref{sec:syner}. 
$A_L$ was determined in~\cite{Gericke2020}
~to be $A_L =(1.55 \pm 0.97 ({\rm stat}) \pm 0.24 ({\rm sys})) \times 10^{-8}$, but was only interpreted in
terms of the DDH parameters in that paper. Using the central values for the updated analysis of the weak couplings given in Table~\ref{tab:Ci}
and setting the vertex function to one,
we obtain $A_L= 2.9 \times 10^{-8}$,
which agrees with the experimental value with uncertainties. Also, we can predict $\frac{d\phi}{dz}(\vec np) = 1.2\times 10^{-6}$ and
$\frac{d\phi}{dz}(\vec nd) = 1.0 \times 10^{-7}$ using the central values only. Using the the vertex function Eq.~\eqref{DDHcutoff} and $\Lambda_\rho = \Lambda_\omega =1.5\,$GeV, one has $\frac{d\phi}{dz}(\vec np) = 0.8\times 10^{-6}$, $\frac{d\phi}{dz}(\vec nd) = 2.7 \times 10^{-7}$
and $A_L= 1.3 \times 10^{-9}$. Clearly, it would be preferable to have sufficiently many data to determine the $C_i$ (together with $h_\pi^1$) independently from any modeling, also 
for further insight into the underlying 
dynamics of these processes.

\subsection{Parity violation in nuclear systems} \label{sec:APV}

Parity violation has been studied theoretically in few-nucleon systems and light- and medium-mass nuclei using variety of methods.
The work based on chiral EFT reported in Sec.~\ref{sec:chiral_few} is not repeated here.\\

\noindent \underline{Few Nucleon Systems}:\\

Considering the neutron-proton system, parity-violating observables such as the longitudinal asymmetry and
neutron-spin rotation in np elastic scattering, the photon asymmetry in $\vec{n}p$ radiative capture, and the asymmetries in deuteron photodisintegration $d(\vec\gamma,n)p$ in the threshold region and electrodisintegration $d(\vec{e},e^\prime)np$ in quasielastic kinematics have been calculated using parity-conserving precision NN
interactions and the parity-violating DDH interaction in 
\cite{Schiavilla2004}. It was found that the latter process provides a very clean probe of the electroweak properties of individual nucleons.

In a related study, the parity-violating longitudinal asymmetry has been studied in the $pp$ elastic scattering~\cite{Carlson2002}. Using several parity-conserving strong-interaction potentials and the parity-violating DDH interaction, the scattering problem was solved in both configuration and momentum space. The predicted parity-violating asymmetries were found to be only weakly dependent upon the input strong-interaction potential. Values for the $\rho$ and $\omega$-meson weak coupling constants $h_\rho^{pp}$ and $h_\omega^{pp}$ were determined by reproducing the measured asymmetries at energies in the range of 13 to 220 MeV. Earlier,
weak couplings from the topological soliton model combined with the
Bonn potential had been used to predict the asymmetry for the TRIUMF experiment~\cite{Driscoll:1989jv}.

In a three-nucleon system, the neutron spin rotation induced by parity-violating components in the NN potential has been studied in the $\vec{n}-d$ scattering of polarized neutrons on the deuteron at zero energy~\cite{Schiavilla2008}. Using the Argonne $v_{18}$ NN and Urbana-IX three-nucleon (3N) parity-conserving interaction in combination with the DDH parity-violating NN potential, it has been found that this observable is dominated by the long-range part of the parity-violating NN potential associated with the pion exchange. Consequently, its measurement could provide a further constraint, complementary to that coming from measurements of the photon asymmetry in $\vec{n}p$ radiative capture~\cite{Schiavilla2004}, on the strength of this component of the hadronic weak interaction. On the technical side, the three-nucleon scattering calculations~\cite{Schiavilla2008} have been performed in the configuration space using Monte Carlo integration for the spatial integrals.

Moving on to the four-nucleon systems, the longitudinal asymmetry induced by parity-violating components in the NN potential has been studied in the charge-exchange reaction $^3$He($\vec{n},p$)$^3$H at vanishing incident neutron energies~\cite{Viviani2010}. The coupled-channel calculations have been performed using the hyperspherical harmonics formalism considering $J{=}0$ and $J{=}1$ $S$ waves in the incoming $n$-$^3$He and outgoing $p$-$^3$H channels with the parity-violating transitions obtained in the first-order perturbation theory in the hadronic weak-interaction potential. The chiral NN and 3N parity-conserving interactions were used in combination with the parity-violating DDH or pionless EFT weak potentials. A rather large range of asymmetries from $\sim -9\times10^{-8}$ to $~\sim -2\times10^{-8}$ has been obtained depending on the input strong-interaction Hamiltonian. This large model dependence is a consequence of cancellations between long-range (pion) and short-range (vector-meson) contributions and is sensitive to the assumed values for the PV coupling constants.

Parity-violating neutron spin rotation has also been investigated in scattering on $^4$He at vanishing incident neutron energy limit. \textit{Ab initio} calculations within the Faddeev-Yakubovsky formalism in configuration space have been reported in 
\cite{Lazauskas2019}. Modern strong-interaction Hamiltonian based on chiral NN+3N interaction and the parity-violating DDH NN interaction was used. The study also discussed an implication of the theoretical large-$N_c$ estimation of weak couplings. The obtained $n-^4$He spin rotation results were in line with the current experimental bounds.\\

\noindent \underline{Atomic Systems}:\\

Precision calculations of parity-violating observables in few-nucleon $A=2-5$ systems combined with accurate measurements will allow us to carry out a systematic program to determine low-energy constants of PV NN interaction. Knowledge of these constants will promote theoretical parity-violation studies to the level of quantitative predictions also in light nuclei and beyond.

In atomic nuclei, the PV NN interaction admixes opposite parity contributions into the ground state. A new static electromagnetic moment of the nucleus can then arise, the so-called anapole moment. It is a parity-odd, time-reversal-even $E1$ moment of the electromagnetic current operator~\cite{Haxton2002}. Although the existence of this moment was recognized theoretically soon after the discovery of parity nonconservation, its experimental observation was achieved only recently in a measurement of the hyperfine dependence of atomic parity nonconservation in $^{133}$Cs~\cite{Wood1997}.

Atomic parity violation (APV) provides a complementary, leptonic probe of HPV at very low momentum transfer. Unlike purely hadronic measurements—where parity-violating amplitudes must be isolated against large strong-interaction backgrounds—APV measures weak neutral-current effects in atoms at momentum transfers $q^2\ll M_Z^2$, thereby accessing the same underlying weak interaction that ultimately generates HPV observables, but in a different experimental environment~\cite{Khriplovich1991, Fortson1990}. APV signals decompose into a nuclear-spin-independent (NSI) component, dominated in heavy atoms by the coherent weak charge $Q_W$ and primarily testing the electroweak sector of the SM, 
and a nuclear-spin-dependent (NSD) component, which 
can appear in heavy atoms (for a comprehensive review, see also \cite{Safronova2018}).\\

\noindent \underline{Low-Energy Effective Electron-Quark Hamiltonian}:  \label{sec:APV:theo-framework}\\

In atomic systems, the momentum transfer is negligible compared to the mass of the $Z^0$ boson ($q^2 \ll M_Z^2$). Consequently, the exchange of a $Z^0$ boson can be accurately described by a point-like, four-fermion contact interaction \cite{Khriplovich1991, Ginges2004}. The parity-violating portion of this effective electron-quark Hamiltonian arises from the interference between the vector ($V$) and axial-vector ($A$) currents and is conventionally separated into two distinct terms \cite{Khriplovich1991, Erler2005}: \footnote{While the $u$ and $d$ quarks dominate the interaction with the nucleus, precision treatments may include possible strange-quark contributions to nucleon axial and vector form factors (note, e.g., \cite{SAMPLE:2000ptk}); their impact on APV is generally subleading at current precision.}

\begin{equation}
H_{\mathrm{PV}} = \frac{G_F}{\sqrt{2}} \sum_{q=u,d,s,...} \left( C_{1q}\, \bar{e}\gamma_\mu\gamma_5 e\, \bar{q}\gamma^\mu q + C_{2q}\, \bar{e}\gamma_\mu e\, \bar{q}\gamma^\mu\gamma_5 q \right)
\label{eq:HPV}
\end{equation}
Here we employ the notation and conventions of 
Sec.~\ref{sec:th-SM}, and the 
two sets of dimensionless coupling constants govern the two distinct manifestations of parity violation in the atom.

\textbf{The Nuclear Spin-Independent (NSI) Term ($C_{1q}$)}---The first term, characterized by $C_{1q}$, couples the axial-vector electron current ($A_e$) to the vector quark current ($V_q$). Because the vector current is coherent over the constituent quarks, this term yields a weak charge $Q_W$ dominated by neutron number $N$. In the tree-level SM, 
$Q_W = -N + Z(1-4\sin^2\theta_W)$ with the atomic number $Z$ and the weak mixing angle $\theta_W$; $Q_W$ primarily scales with $N$. The proton contribution is numerically small but not negligible (e.g., for Cs it is $\sim 3.5\%$ of the total \cite{Cadeddu2019}). The approximate $\sim Z^3$ enhancement arises from two factors: (i) nuclear coherence in the vector quark current, giving $\sim N$ (with $N \sim Z$ for stable isotopes); and (ii) relativistic enhancement of the electron wavefunction at the nucleus, $|\psi(0)|^2$, scaling as $\sim Z^{2+\epsilon}$ for large $Z$. \cite{Bouchiat1974} The precise scaling is isotope-dependent (e.g., $N/Z\approx 1.4$ in $^{133}$Cs). Early experimental confirmation came from Barkov \& Zolotorev \cite{BarkovZolotorev1978} in bismuth and Bouchiat \& Pottier \cite{Bouchiat1985} in cesium.

The couplings $C_{1u}$ and $C_{1d}$ combine to define the weak charge of the nucleus \cite{Blundell1990, Khriplovich1991, Erler2005}. For cesium specifically, high-precision many-body perturbation theory calculations by Dzuba, Flambaum, Silvestrov, \& Sushkov \cite{Dzuba1985} established the theoretical framework for extracting weak charges, with subsequent refinements \cite{Porsev2009, Dzuba2012} addressing correlation effects crucial for sub-percent precision. Including electroweak radiative corrections, the effective weak charge can be written schematically as:
\begin{equation}
Q_W \approx \rho_{\rm NC} \left[ -N + Z(1 - 4\sin^2\theta_W) \right] + \text{box corrections}
\end{equation} 
where $\rho_{\rm NC}$ is a multiplicative radiative correction, and ``box corrections'' represent additional contributions from $\gamma Z$ box diagrams \cite{Marciano1983}. Updated electroweak radiative corrections specific to APV were later provided by Czarnecki \& Marciano \cite{CzarneckiMarciano1996}, while QED corrections in heavy atoms were comprehensively treated by Blundell, Johnson, \& Sapirstein \cite{Blundell1990}. In precision work, these corrections are typically applied separately rather than absorbed into a single parameter. In heavy atoms like cesium or radium, this coherent scaling dictates that the NSI term yields the dominant contribution to the overall APV signal \cite{Wood1997, Porsev2009}. While this provides a highly precise test of the SM 
electroweak sector, it contains minimal information regarding internal hadronic dynamics.

\textbf{The Nuclear Spin-Dependent (NSD) Term ($C_{2q}$)}---The second term, governed by the quark-level couplings $C_{2q}$, couples the vector electron current ($V_e$) to the axial-vector quark current ($A_q$). Unlike the NSI term, the axial-vector quark current depends on the spin of the quarks. When evaluated at the nuclear level, this interaction is parameterized by the effective nucleon couplings $C_{2N}$. Because the spins of paired nucleons effectively cancel, this term lacks atomic mass scaling and relies primarily on the spin of the unpaired valence nucleon \cite{Ginges2004}. The couplings $C_{2q}$ are proportional to the product of the electron's vector coupling $g_V^e \propto (1-4\sin^2\theta_W)$ and the quark's axial coupling $g_A^q$. Because $1-4\sin^2\theta_W \approx 0.04$ at tree level, all $C_{2q}$ are numerically small. Translated to the nucleon level, the effective couplings $C_{2p}$ and $C_{2n}$ are related to combinations of $C_{2u}$ and $C_{2d}$ weighted by the nucleon's quark content; explicitly,
\begin{equation}
C_{\text{2p}} = -C_{\text{2n}} = g_A(1-4\sin^2\theta_W)/2 \simeq 0.05,
\end{equation}
with $g_A \simeq 1.27$ and 
sin$^2\theta_W \simeq 0.23$~\cite{ParticleDataGroup:2024cfk}.
Radiative corrections---including contributions from $\gamma Z$ box diagrams---can modify these couplings significantly \cite{Marciano1983, Erler2013}. 

\textbf{Sensitivity to new neutral gauge bosons}---New neutral gauge bosons ($Z'$) arise naturally in many extensions of the 
SM~\cite{Langacker2009}, including grand unified theories, left-right symmetric models, and string-inspired constructions, and their discovery would provide direct evidence for an extended gauge sector --- which can include lighter mass gauge bosons of a ``dark sector'' as well~\cite{Bouchiat:2004sp,Davoudiasl:2012ag}. 
APV measurements can probe $Z'$ bosons through their virtual contributions to the low-energy electron-quark parity-violating couplings, complementing direct searches at high-energy colliders \cite{Buckley2012, Safronova2018}. In the NSD sector specifically, $Z'$ exchange shifts the effective couplings $C_{2q}$---which parameterize the interaction between the vector electron current and the axial-vector quark current---by an amount proportional to $g_V^e \times g_A^q / M_{Z'}^2$. Since the $C_{2q}$ couplings are accidentally suppressed in the 
SM by the proximity of $\sin^2\theta_W$ to $1/4$, they are exceptionally sensitive to such new physics contributions, and the combination of $Z'$ couplings they probe ($g_V^e \times g_A^q$) is complementary to the $g_A^e \times g_V^q$ combination accessed by nuclear-spin-independent measurements such as the nuclear weak charge \cite{Marciano1990}.\\

\noindent \underline{The Nuclear Anapole Moment}:\\

To understand how hadronic parity violation dominates the nuclear spin-dependent (NSD) atomic signal, one must examine the electromagnetic structure of the nucleus. In 1958, Y. B. Zel'dovich (and independently V.G. Vaks) demonstrated that a system lacking parity ($P$) symmetry but strictly retaining time-reversal ($T$) symmetry can possess a unique electromagnetic multipole: the anapole moment \cite{Zeldovich1958}. Classically, the anapole moment corresponds to the magnetic field generated by a toroidal electric current. The magnetic field is entirely confined within the torus, meaning it only interacts with particles that physically penetrate the current distribution.

In a nucleus, this parity-violating structure arises from weak interactions between nucleons. While the strong nuclear force binds the protons and neutrons together in states of definite parity, the weak interaction manifests at the hadronic scale as a parity-violating NN potential, $V_{\text{PNC}}$, which is commonly described 
in a one-boson exchange
model, such as that of DDH\cite{DDH1980, Adelberger1985}, or in 
ChEFT~\cite{Zhu2005}. This potential weakly admixes nuclear states of
opposite parity~\cite{DDH1980, Zhu2005, Adelberger1985, deVries2013,Vanasse2012}.
(ChEFT is based on the symmetries of the SM, 
so that its unknown low-energy constants
could potentially be expressed in terms of
fundamental SM parameters.) 
This parity admixture induces an asymmetric, chiral current density $\mathbf{J}(\mathbf{r})$ within the nucleus. The resulting nuclear anapole moment vector, $\mathbf{a}$, is defined as the second moment of this internal current distribution \cite{Haxton2001APV}:
\begin{equation}
    \mathbf{a} = -\pi \int d^3r \, r^2 \mathbf{J}(\mathbf{r})
\end{equation}
\textit{(Note: We follow the convention used in \cite{Haxton2001APV}, with $\hbar=c=1$ and the normalization chosen such that the electron–anapole Hamiltonian below is written in terms of the dimensionless $\kappa_a$.)} Microscopic nuclear-structure calculations needed to evaluate this integral for specific nuclei were developed in early nuclear anapole-moment studies (e.g., \cite{Flambaum1997}, \cite{Haxton2001PRL}), with subsequent refinements by Liu \cite{Liu2007} that revisit the calculation of parity-violating two-body exchange currents.

Unlike an EDM, 
which violates both $P$ and $T$ symmetries, the anapole moment is \(P\)-odd and \(T\)-even. In the absence of external fields, rotational invariance implies that the expectation value of the anapole moment in a nucleus with total spin \(\mathbf{I}\) must be aligned with \(\mathbf{I}\). It is therefore parameterized as $\mathbf{a} = a_0 \kappa_a \mathbf{I}$, where $a_0$ is a normalization constant (setting dimension and units), and $\kappa_a$ is the dimensionless anapole moment parameter encoding the strength of the hadronic PV admixture \cite{Khriplovich1991}.

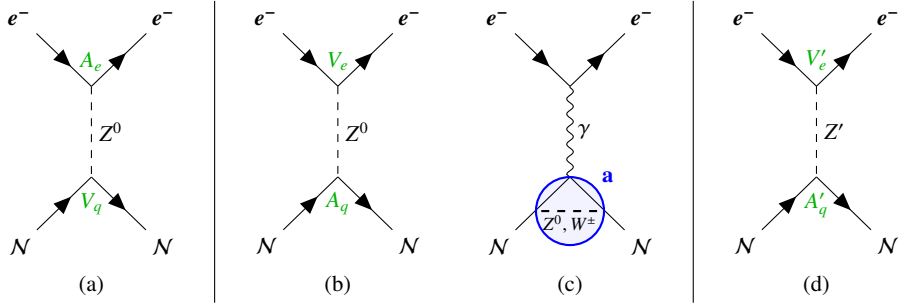
\begin{figure}[ht!]
    \centering
    \begin{minipage}{0.15\textwidth}
        \centering
        \begin{tikzpicture}
            \begin{feynman}
                \vertex (v1);
                \vertex [below=1.2cm of v1] (v2);
                \vertex [above left=0.7cm and 0.7cm of v1] (t1) {\(\boldsymbol{e^{-}}\)};
                \vertex [above right=0.7cm and 0.7cm of v1] (t2) {\(\boldsymbol{e^{-}}\)};
                \vertex [below left=0.7cm and 0.7cm of v2] (b1) {\(\mathcal{N}\)};
                \vertex [below right=0.7cm and 0.7cm of v2] (b2) {\(\mathcal{N}\)};
                
                \diagram* {
                    (t1) -- [fermion] (v1) -- [fermion] (t2),
                    (v1) -- [scalar, edge label=\(Z^0\)] (v2),
                    (b1) -- [fermion] (v2) -- [fermion] (b2),
                };
                
                \node[green!70!black, above=0.1cm of v1] {\(A_e\)};
                \node[green!70!black, below=0.1cm of v2] {\(V_q\)};
                \node[black, below=1.2cm of v2] {(a)};
            \end{feynman}
        \end{tikzpicture}
    \end{minipage}%
    \hspace{0.3cm} \vrule \hspace{0.3cm}
    \begin{minipage}{0.15\textwidth}
        \centering
        \begin{tikzpicture}
            \begin{feynman}
                \vertex (v1);
                \vertex [below=1.2cm of v1] (v2);
                \vertex [above left=0.7cm and 0.7cm of v1] (t1) {\(\boldsymbol{e^{-}}\)};
                \vertex [above right=0.7cm and 0.7cm of v1] (t2) {\(\boldsymbol{e^{-}}\)};
                \vertex [below left=0.7cm and 0.7cm of v2] (b1) {\(\mathcal{N}\)};
                \vertex [below right=0.7cm and 0.7cm of v2] (b2) {\(\mathcal{N}\)};
                
                \diagram* {
                    (t1) -- [fermion] (v1) -- [fermion] (t2),
                    (v1) -- [scalar, edge label=\(Z^0\)] (v2),
                    (b1) -- [fermion] (v2) -- [fermion] (b2),
                };
                
                \node[green!70!black, above=0.1cm of v1] {\(V_e\)};
                \node[green!70!black, below=0.1cm of v2] {\(A_q\)};
                \node[black, below=1.2cm of v2] {(b)};
            \end{feynman}
        \end{tikzpicture}
    \end{minipage}%
    \hspace{0.5cm}
    \begin{minipage}{0.15\textwidth}
        \centering
        \begin{tikzpicture}
            \begin{feynman}
                \vertex (v1);
                \vertex [below=1.2cm of v1] (v2);
                \vertex [above left=0.7cm and 0.7cm of v1] (t1) {\(\boldsymbol{e^{-}}\)};
                \vertex [above right=0.7cm and 0.7cm of v1] (t2) {\(\boldsymbol{e^{-}}\)};
                \vertex [below left=0.7cm and 0.7cm of v2] (b1) {\(\mathcal{N}\)};
                \vertex [below right=0.7cm and 0.7cm of v2] (b2) {\(\mathcal{N}\)};
                \node[black, below=1.2cm of v2] {(c)};
                
                \diagram* {
                    (t1) -- [fermion] (v1) -- [fermion] (t2),
                    (v1) -- [photon, edge label=\(\gamma\)] (v2),
                    (b1) -- [plain] (v2) -- [plain] (b2),
                };
            \end{feynman}
            
            \coordinate (c_center) at ([yshift=-0.45cm]v2);
            \filldraw[blue, thick, fill=blue, fill opacity=0.05] (c_center) circle (0.45cm);
            \draw[black, dashed, thick] ([xshift=-0.38cm]c_center) -- ([xshift=0.38cm]c_center);
            \node[black, font=\scriptsize] at ([yshift=-0.17cm]c_center) {\( Z^0, W^{\pm} \)};
            \node[blue, font=\normalsize] at ([xshift=0.5cm]v2) {$\mathbf{a}$};
            
        \end{tikzpicture}
    \end{minipage}%
    \hspace{0.3cm} \vrule \hspace{0.3cm}
    \begin{minipage}{0.15\textwidth}
        \centering
        \begin{tikzpicture}
            \begin{feynman}
                \vertex (v1);
                \vertex [below=1.2cm of v1] (v2);
                \vertex [above left=0.7cm and 0.7cm of v1] (t1) {\(\boldsymbol{e^{-}}\)};
                \vertex [above right=0.7cm and 0.7cm of v1] (t2) {\(\boldsymbol{e^{-}}\)};
                \vertex [below left=0.7cm and 0.7cm of v2] (b1) {\(\mathcal{N}\)};
                \vertex [below right=0.7cm and 0.7cm of v2] (b2) {\(\mathcal{N}\)};
                \node[black, below=1.2cm of v2] {(d)};
                
                \diagram* {
                    (t1) -- [fermion] (v1) -- [fermion] (t2),
                    (v1) -- [scalar, edge label=\(Z'\)] (v2),
                    (b1) -- [fermion] (v2) -- [fermion] (b2),
                };
                
                \node[green!70!black, above=0.1cm of v1] {\(V_e'\)};
                \node[green!70!black, below=0.1cm of v2] {\(A_q'\)};
            \end{feynman}
        \end{tikzpicture}
    \end{minipage}
\caption{Diagrams contributing to atomic parity violation. (a) Nuclear-spin-independent (NSI) contribution from $Z^0$ exchange ($A_e \times V_q$). (b)--(d) Nuclear-spin-dependent (NSD) contributions: (b) tree-level $Z^0$ exchange ($V_e \times A_q$), (c) nuclear anapole moment $\mathbf{a}$, (d) beyond the SM 
tree-level Z' exchange ($V'_e \times A'_q$). Note: $Z'$ also contributes to NSI via the ($A'_e \times V'_q$) combination.}
\label{fig:apv-diagrams}
\end{figure}

Because the anapole magnetic field is confined within the nuclear volume, an atomic electron only interacts with it if its wave function exhibits non-zero overlap with the nucleus. The effective Hamiltonian describing this interaction is \cite{Ginges2004}:
\begin{equation}\label{eq:anap}
H_{\text{anapole}} = \frac{G_F}{\sqrt{2}} \frac{K}{I(I+1)} \kappa_a \mathbf{\alpha} \cdot \mathbf{I} \rho(r),
\end{equation}
where $K \equiv (I+1/2)(-1)^{(I+1/2-l)}$ is the standard angular factor ($l$ is valence nucleon orbital angular momentum), $I$ is nuclear spin, $\mathbf{I}$ is the nuclear spin operator, $\mathbf{\alpha}$ is the Dirac matrix vector, and $\rho(r)$ is the normalized nuclear density. In an APV experiment, the anapole moment cannot be measured in pure isolation. The experimentally extracted NSD parameter, $\kappa_{\text{total}}$, is a linear combination of several overlapping effects:
\begin{equation}\label{eq:kappa_total}
\kappa_{\text{total}} = \kappa_a + \kappa_Z + \kappa_{\text{hf}}~.
\end{equation}
Here, $\kappa_a$ is the anapole moment contribution \cite{Flambaum1984} (see also Fig.~\ref{fig:apv-diagrams}~(c)), $\kappa_Z$ represents the tree-level electroweak $Z^0$ exchange between the electron and the unpaired nucleon \cite{Flambaum1980} (see also Fig.~\ref{fig:apv-diagrams}~(b)), and $\kappa_{\text{hf}}$ accounts for hyperfine-induced mixing of the much larger NSI weak charge \cite{Bouchiat1991, Johnson2003}. Beyond the 
SM contributions could also be present; see Fig.~\ref{fig:apv-diagrams}~(d) and the discussion of $Z'$ sensitivity above.

While both the anapole and $Z^0$ exchange interactions share the same mathematical form ($H \propto \boldsymbol{\alpha} \cdot \mathbf{I}$) and benefit equally from the $Z^2$ enhancement of the atomic electron wavefunction at the nucleus, they scale differently with respect to nuclear mass. Flambaum and Khriplovich estimated that the nuclear anapole moment experiences a coherent nuclear enhancement scaling roughly as $A^{2/3}$ \cite{Flambaum1980}. This estimate is intended as a parametric guide; quantitative values depend strongly on shell structure, core polarization, and two-body current contributions \cite{Flambaum1980, Haxton2001APV}. Because the tree-level $Z^0$ exchange ($\kappa_Z$) involves only the unpaired valence nucleon and carries no collective nuclear enhancement, the anapole contribution \(\kappa_a\) typically dominates \(\kappa_{\text{total}}\) in heavy atoms, although the relative sizes of \(\kappa_a\), \(\kappa_Z\), and \(\kappa_{\mathrm{hf}}\) remain nucleus- and atom-dependent and must be evaluated case by case \cite{Haxton2001APV, Ginges2004}. In light atoms, by contrast, the anapole and $Z^0$-exchange contributions can be comparable, making nuclear spin-dependent parity violation measurements in light nuclei sensitive tests of the SM 
and probes of new particles such as $Z'$ bosons and particles contributing to electroweak radiative corrections. Consequently, isolating $\kappa_a$ from atomic data—relying on robust atomic many-body theory to subtract $\kappa_{\text{hf}}$ and $\kappa_Z$—provides a direct, low-energy window into hadronic parity violation.\\

\noindent \underline{Many-body computation of the anapole moment}:\label{sec:many-body}\\

The nuclear anapole moment can be calculated by evaluating the anapole operator mean value in the ground-state wave function of the nucleus with the opposite parity admixture included. As the parity violating NN interaction is orders of magnitude weaker than the parity-conserving nuclear force, one can apply the second-order perturbation theory to obtain the full wave function, i.e.,
\begin{equation}\label{gswf}
  |\psi_{\rm gs}\; I \rangle = |\psi_{\rm gs}\; I^\pi \rangle + \sum_j  |\psi_j \; I^{-\pi}\rangle \;
                                 \frac{1}{E_{\rm gs}-E_j} \; \langle \psi_j \; I^{-\pi}| V_{\rm NN}^{\rm PV}|\psi_{\rm gs} \; I^\pi \rangle \; .
\end{equation}  
Here, $I$ is the ground-state total angular momentum, $|\psi_{\rm gs}\; I^\pi \rangle$ is the ground-state wave function obtained by solving the Schr\''{o}dinger equation with the parity-conserving Hamiltonian $H_{\rm PC}$ with strong and electromagnetic NN (and 3N) interactions, $\pi$ is the ground-state (natural) parity obtained from the PC calculation. The $V_{\rm NN}^{\rm PV}$  interaction is the DDH or the chiral EFT PV NN interaction. The sum in Eq.~(\ref{gswf}) includes all states of the angular momentum $I$ and the unnatural parity coupled by the PV NN interaction, which might be intractable to evaluate explicitly. However, when applying configuration-interaction methods such as nuclear shell model or the \textit{ab initio} no-core shell model (NCSM)~\cite{Barrett2013},  it is not necessary to compute many excited unnatural parity states as Eq.~(\ref{gswf}) suggests. Rather, the wave function $|\psi_{\rm gs}\; I \rangle$ can be obtained by solving the Schr\"{o}dinger equation with an inhomogeneous term
\begin{equation}\label{inhomeq}
  (E_{\rm gs}-H_{\rm PC})  |\psi_{\rm gs}\; I \rangle =  V_{\rm NN}^{\rm PV}|\psi_{\rm gs} \; I^\pi \rangle \; .
\end{equation}
To invert this equation and obtain the ground-state wave function with the unnatural parity admixed, one can apply the Lanczos continued fraction algorithm~\cite{Lanczos1950,Haydock1974}.

The leading contribution to the anapole moment operator is given by the spin term~\cite{Flambaum1997},
\begin{equation}\label{as}
  \hat{a}_s=\frac{\pi e}{m} \sum_{i=1}^A \mu_i (\vec{r}_i\times\vec{\sigma}_i) \; ,
\end{equation}
with $m$ the nucleon mass and $\mu_i$ the nucleon magnetic moment in units of nuclear magneton, i.e., $\mu_i{=}\mu_p (1/2{+}t_{z,i})+\mu_n(1/2{-}t_{z,i})$ with $\mu_p{=}2.79$ and $\mu_n{=}-1.91$. The anapole moment is then evaluated using
\begin{equation}\label{anapole}
a_s=\langle \psi_{\rm gs}\; I \; I_a=I | ~~\hat{a}^{(1)}_{s,0}~| \psi_{\rm gs}\; I \; I_a=I\rangle \; ,
\end{equation}
and typically expressed in terms of a dimensionless coupling constant $\kappa_a$, see Eqs.~(\ref{eq:anap}) and (\ref{eq:kappa_total}).

Calculations of the anapole moment of the deuteron using the PC AV18 NN potential and the meson-exchange based PV NN interactions have been reported in \cite{Hyun2023,Liu2003}. It has been observed that the contribution of heavy mesons ($\rho, \omega$) was suppressed by two orders of magnitude compared to the pion one.

Anapole moments of light stable isotopes $^9$Be, $^{13}$C, $^{14,15}$N, $^{25}$Mg have been calculated within the \textit{ab initio} NCSM in the context of planned experiments to investigate nuclear spin-dependent parity-violating effects in tri-atomic molecules~\cite{Hao2020}. The NCSM is a configuration interaction basis-expansion many-body method that uses harmonic-oscillator basis. The calculations employed precision chiral EFT PC NN+3N interaction combined with the DDH PV NN interaction and demonstrated a good convergence with the basis size. Obtained anapole moments were typically a factor of 2 to 3 larger in absolute value than the single-particle estimates~\cite{Hao2020}.

One of the light nuclei of experimental interest regarding the anapole moment measurement is $^{19}$F. The NCSM can be applied to investigate parity-violation in this isotope as demonstrated in calculations of its nuclear Schiff moment~\cite{Ng_2026}. The $^{19}$F anapole moment calculations within NCSM are in progress. 

Anapole and other parity-violating moments of medium-mass and heavy nuclei can be investigated by applying methods that scale polynomial with the number of nucleons $A$ such as the in-medium similarity renormalization group (IMSRG)~\cite{Hergert2016,Hergert2017} or the coupled-cluster method (CCM)~\cite{Hagen_2014}. The valence-space variant of the IMSRG approach, VS-IMSRG, is particularly promising with the same reach as the phenomenological nuclear shell model.

The IMSRG employs a continuous unitary transformation to reshape the Hamiltonian into a form that is easier to diagonalize. 
This is achieved through the flow equation,
\begin{equation}\label{eq:flow}
    \frac{dH(s)}{ds} = \left[\eta(s), H(s)\right],
\end{equation}
where $H(s)$ is the Hamiltonian as a function of the flow parameter $s$, and $\eta(s)$, the generator of the transformation, encodes the degrees of freedom one wishes to decouple. 
Any other operator $O$ must be evolved consistently under the same
transformation. Until recently, the IMSRG has been restricted to symmetry-conserving observables. The computation of parity-violating moments requires access to excited states outside of the valence shell as seen in Eq.~(\ref{gswf}), which would require complicated extensions of the method. However, following the ideas from the density-functional theory~\cite{Ban2010}, it has been proposed to treat the effects of parity-violating interaction non-perturbatively, on the same footing as the strong nuclear Hamiltonian~\cite{Engel2025,Belley2026}. By evolving the strong Hamiltonian and symmetry-breaking potential together through a continuous unitary transformation, one can consistently decouple the low-energy dynamics and compute parity-violating observables without requiring explicit sums over excited states.

In this framework, one  starts from a Hamiltonian that includes both parity-conserving ($H_{\rm PC}$) and parity-violating ($V_{\rm NN}^{\rm PV}$) interactions,
\begin{equation}
    H = H_{\rm PC} + \lambda V_{\rm NN}^{\rm PV},\label{eq:H}
\end{equation}
where $\lambda$ is a power-counting parameter used to track small terms that one ultimately sets to 1 (as is typical in perturbation theory). The corresponding generator can be written as
\begin{equation}
    \eta = \eta_{\rm PC} + \lambda \eta_{\rm PV}.\label{eq:eta}
\end{equation}
Inserting these definitions into Eq.~\eqref{eq:flow}, one obtains the coupled flow equations
\begin{align}
    \frac{dH_{\rm PC}(s)}{ds} &= \left[\eta_{\rm PC}(s), H_{\rm PC}(s)\right] 
    + \lambda^2 \left[\eta_{\rm PV}(s), V_{\rm NN}^{\rm PV}(s)\right],\label{eq:HPC_evol} \\
   \frac{dV_{\rm NN}^{\rm PV}(s)}{ds} &= \left[\eta_{\rm PC}(s), V_{\rm NN}^{\rm PV}(s)\right]+ \left[\eta_{\rm PV}(s), H_{\rm PC}(s)\right]\label{eq:HPV_evol}.
\end{align}
For a generic operator $O = O_{\rm PC} + O_{\rm PV}$ consisting of parity-conserving and parity-violating parts, one obtains analogous equations~\cite{Belley2026}.
A key feature of these equations is that even an operator that initially has negative parity, so that $O_{\rm PC}(0)=0$, will acquire a parity-conserving part that represents, the effects of opposite-parity virtual excitations.   

This approach has been successfully implemented and benchmarked with the NCSM calculations discussed above~\cite{Belley2026}. The first results for a nucleus of experimental interest, $^{29}$Si, have been reported in the same work. Applications to other medium-mass nuclei of experimental interest such as $^{133}$Cs are in progress.

The $Z^0$-exchange contribution to the NSD parity-violating signal — i.e., the $\kappa_Z$ piece of Eq. (16) — is straightforward to evaluate in nuclear many-body methods. Using $C_{2p} = -C_{2n} \simeq 0.05$ from Eq. (11), one has $\kappa_\text{ax} {=}\, I\, \kappa_Z {=} -2 C_{2p} \langle s_{p,z} \rangle - 2 C_{2n} \langle s_{n,z} \rangle {\simeq}-0.1 \langle s_{p,z}\rangle{+}0.1 \langle s_{n,z}\rangle$~\cite{Hao2020}, where the spin operator matrix elements are
\begin{eqnarray}
    \langle s_{\nu,z} \rangle{\equiv}\langle \psi_{\text{gs}} \; I^\pi I_z{=}I|s_{\nu,z}|\psi_{\text{gs}} \; I^\pi I_z{=}I\rangle, 
\end{eqnarray}
with $\nu{\equiv} p$ or $n$.

\begin{figure}[htb!]
\centering\includegraphics[scale=0.3]{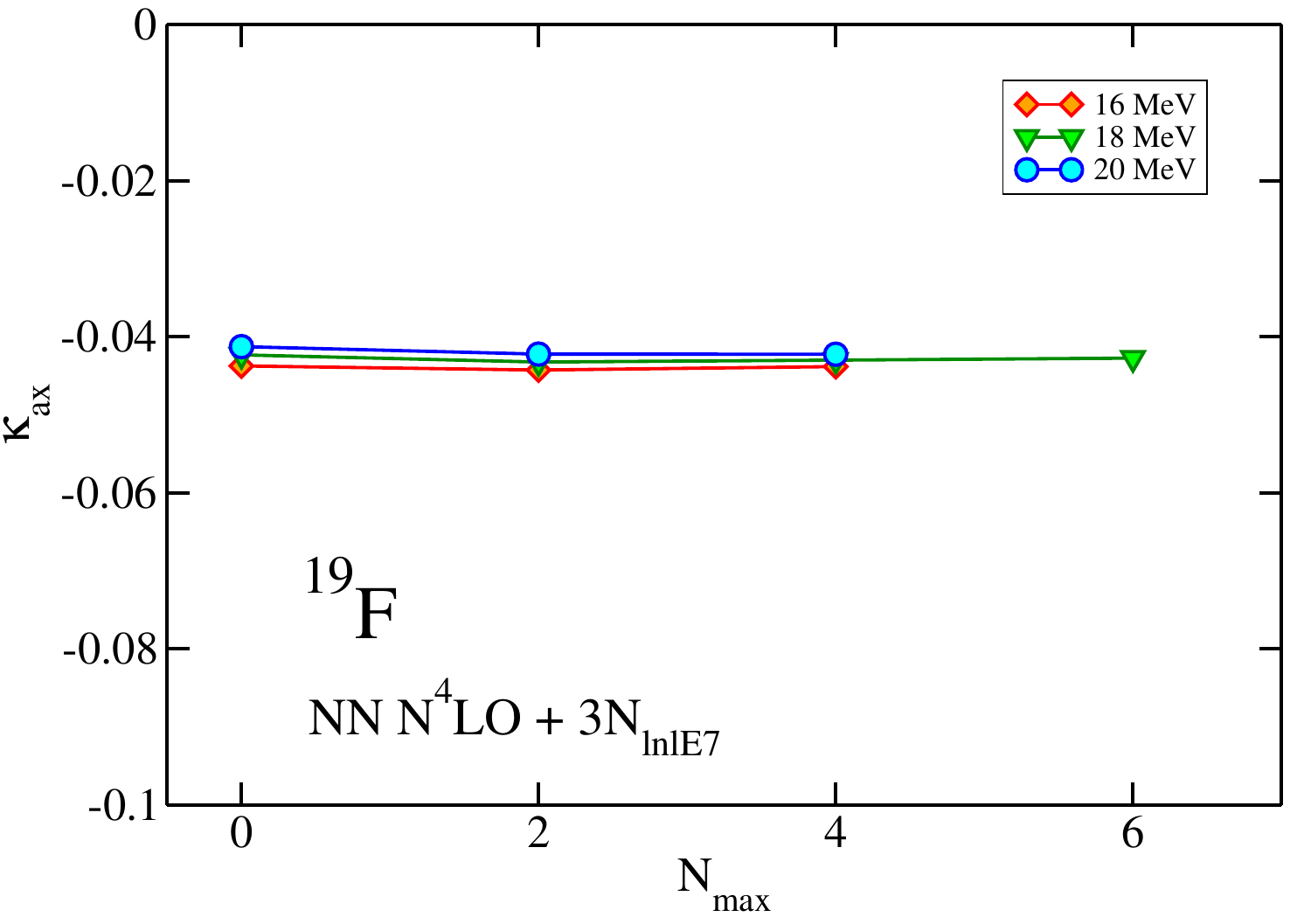}
\caption{Coupling constant $\kappa_{ax} {=}\, I\, \kappa_Z $ for $^{19}$F ($I{=}1/2$) calculated within NCSM. Good convergence is found with respect to the basis size characterized by $N_{\rm max}$ and the harmonic oscillator frequency $\hbar\Omega$ shown in the legend. The chiral NN+3N interaction and wave functions are as described in \cite{Ng_2026}.}
\label{fig:19F_kappaax}
\end{figure}

These are straightforward to compute in ab initio many-body methods or in the phenomenological nuclear shell model. In the NCSM they manifest excellent basis-size convergence, as illustrated for $^{19}$F in Fig.~\ref{fig:19F_kappaax}.

Calculations in light nuclei~\cite{Hao2020} confirm that $\kappa_a$ and $\kappa_{ax}$ are of comparable size. Interestingly, all reported results there show 
coherent contributions from the two processes. The same was found in $^{19}$F calculations using wave functions from \cite{Ng_2026}.

Anapole moments of $^{133}$Cs and $^{205}$Tl for which experiments were performed, have been investigated together with the spin-dependent PV contribution due to the Z-boson exchange within the nuclear shell model using configuration-mixed shell-model wave functions and the DDH PV NN interaction~\cite{Haxton2002}. The experimental anapole moment constraints on the PV meson-nucleon coupling constants were derived and compared with those from other tests of the hadronic weak interaction. It has been found that while the bounds obtained from the anapole moment results were consistent with the broad ‘‘reasonable ranges’’ in 
coupling constants 
suggested by DDH~\cite{DDH1980}, 
they were not in good agreement with the constraints from the other experiments, 
as we have noted in earlier sections. 
This outcome is reflected in Fig. 9 in \cite{Haxton2002} and elsewhere. 
An inadequate description of the nuclear structure of these complex nuclei has been pointed out as the most likely reason for the discrepancies. This highlights the importance and urgency of performing new \textit{ab initio} calculations with quantifiable uncertainties such as the IMSRG ones described above employing the latest chiral EFT based PV NN interactions to clarify the situation. Despite this, 
we note that anapole moment extractions do 
require additional theoretical and experimental inputs
from atomic physics, which can impact the final 
values of $\kappa_a$ to be interpreted by nuclear theory,  
and we describe this context 
in Sec.~\ref{sec:apv-anapole}.

\section{Experimental Overview
\label{sec:expt}} 

 At energies below $\Lambda_{QCD}$ where nucleons are the relevant degrees of freedom, Danilov showed long ago~\cite{Danilov1965} that five independent weak transition amplitudes are present in NN elastic scattering at low enough energies that only $L=0$ and $L=1$ partial waves are important: the $\Delta I = 1$ transition amplitudes between $^3S_1 -{^3P_1}$ and $^1S_0 -{^3P_0}$ partial waves; the $\Delta I = 0$ transition amplitudes between $^3S_1 -{^1P_1}$ and $^1S_0 -{^3P_0}$ partial waves; and the $\Delta I = 2$ transition amplitude between the $^1S_0 -{^3P_0}$ partial waves. Despite several decades of effort, these amplitudes are not yet fixed by experiment. Since the strong interaction conserves parity and the weak interaction violates parity, the measurement of a parity-odd observable provides a clear, background-free experimental signal for the transition amplitudes of interest. However, 
 as we have already noted, dimensional analysis implies that the typical size of the NN weak amplitudes are about $10^{-7}$ of strong interaction amplitudes, so the signal is quite small. Furthermore, our incomplete understanding of the strong interaction at low energy makes it difficult to interpret a parity-odd observable in terms of these transition amplitudes unless the measurement is conducted in a two-nucleon or few-nucleon system. The isolation of a parity-odd observable in a low-energy reaction among two and few nucleon systems requires one to either polarize the initial state or analyze the polarization of the final state to measure a pseudoscalar observable. Simply gaining experimental access to a sufficiently intense source of (in general polarized) species of photons or nucleons to see a  $10^{-7}$ effect is itself a major challenge. Finally, a typical P-odd observable in a few nucleon dynamical process is some linear combination of all of the transition amplitudes with different weights that depend on the observable of interest. This combination of circumstances explains why NN weak interaction amplitudes are still so poorly understood experimentally.

 Parity-odd effects were seen in many measurements in nuclei well before intense polarized nucleon beams became available, but in all of these cases the size of the P-odd observable was amplified well above the 
 expected $10^{-7}$ effect 
 by some feature of the nuclear structure in the system which is typically difficult to calculate from first principles. Often this amplification comes from the presence of nearly-degenerate opposite parity levels, which are mixed by the P-odd interaction. One expects parity-odd observables in nuclei to be generically more challenging to calculate than parity-even observables. Since the parity of the shells in a simple 3D harmonic oscillator alternate in sign, the calculation of parity-odd observables in a transition between different nuclear bound states is sensitive to mixing from states only one oscillator shell away, rather than two oscillator shells away as for the case of a parity-even observable. For this reason many attempts to calculate P-odd effects even in light nuclei in the s-d shell using nuclear models have historically exhibited poor convergence as the model space is expanded. In heavy nuclei, the impact of intruder states from strong spin-orbit splitting that push levels of one parity into a sea of states of the opposite parity are also important to understand. These observations reinforce the importance of measurements in two and few nucleon systems to determine NN weak amplitudes.
 
Experiments with proton 
beams were the first systems to achieve sufficient intensities, polarization, and control of systematic errors to enable sufficiently sensitive NN and few nucleon measurements to constrain NN weak interaction amplitudes. Sensitive parity-odd asymmetries were sought in the measurement of the P-odd longitudinal asymmetry in proton-proton and proton-$^{4}$He scattering at proton beam facilities such as Bonn, PSI, and TRIUMF. The beam energy must be chosen to overcome Coulomb repulsion so that strong interactions dominate: in some cases this constraint places the reaction energy near or beyond the boundary where effective field theory treatments of the dynamics are applicable. $A_{z}$ in proton-proton scattering was measured at Bonn at 13.6 MeV ($A_{z}= [-1.5 \pm 0.5⁢] \times 10^{-7}$) ~\cite{Eversheim:1991tg} and at PSI at 45 MeV ($A_{z}= [-1.50 \pm 0.22] \times 10^{-7}$)~\cite{Kistryn:1987tq}. The heroic TRIUMF measurement of $A_{z}$ in proton-proton scattering, $A_{z}= [0.84 \pm 0.29⁢(stat.) \pm 0.17⁢(sys.)] \times 10^{-7}$~\cite{TRIUMFE497:2001hga} was conducted at a ``magic''  energy of 221 MeV, where the dominant S − P wave mixing contribution to the parity-odd amplitude integrates to zero, thereby enabling one to resolve the contribution from P-D mixing. 

Over the last two decades polarized slow neutron beams have become intense enough to resolve NN weak interaction observables. The meV kinetic energy of the neutrons used for this work places the reaction energies well within the regime of validity of EFT approaches. The electrical neutrality of the slow neutron beam makes it possible to reverse the neutron spin without noticeably changing any other aspect of the neutron beam phase space, which greatly suppresses a large variety of possible systematic errors in the measurement. Sensitive parity-odd asymmetries were sought in neutron-proton capture, neutron-$^{3}$He capture, and in neutron-$^{4}$He scattering. The NPDGamma collaboration measured a parity-odd asymmetry in the emission of the 2.2 MeV gamma ray relative to the incident neutron spin in $\vec{n}+p \to d+\gamma$ of $A^{np}_{\gamma}=[-3.0 \pm 1.4(stat) \pm 0.2(sys)] \times 10^{-8}$ with $2\sigma$ statistical significance~\cite{NPDGamma:2018vhh}. This determines the $\Delta I=1$,  $^3S_1 -{^3P_1}$ component of the weak NN
interaction, dominated in the meson picture by pion exchange.  The $n ^3$He Collaboration placed a stringent constraint on the P-odd correlation between the neutron polarization and the proton emission direction in the $\vec{n}+^{3}$He$\to ^{3}$H$+p$ reaction of $A_{PV}=[1.58 \pm 0.97 (stat) \pm 0.24 (sys)] \times 10^{-8}$~\cite{Gericke2020}, which is the smallest asymmetry of a parity-odd observable in NN and few nucleon systems measured so far. The NSR collaboration placed an upper bound on the parity-odd neutron rotary power in $\vec{n}+^{4}$He of $d\phi/dz=[+2.1 \pm 8.3(stat.) \pm 2.9(sys.)] \times 10^{-7}$ rad/m~\cite{Swanson:2019cld}. All of these measurements were dominated by statistical errors, which opens the way for improved measurements at higher intensity neutron beams. The NPDGamma result is in mild tension with previous data on the  $^{3}S_{1} \to ^{3}P_{1}$ amplitude from measurements in $^{18}$F/$^{19}$F~\cite{Barnes:1978sq,Adelberger_beta2PV:1983zz,Page:1987ak}, with a theory calculation calibrated from first forbidden beta decay data~\cite{Haxton1981}. 
Nonzero parity-odd asymmetries were also resolved in slow neutron experiments conducted at the Institute Laue-Langevin in the correlation coefficients for the triton emission direction relative to the neutron spin in the $\vec{n}+^{6}$Li$\to ^{3}$H$+^{4}$He reaction of $\alpha_{6Li}=−[8.8 \pm 2.1] \times 10^{-8}$, and for the correlation coefficient for $\gamma$ emission direction relative to the neutron spin in $\vec{n}+^{10}$B$\to ^{4}$He$+^{7}$Li$^{*}\to^{7}$Li$+\gamma$ reaction of $\alpha_{10B}=−[11.2 \pm 3.4] \times 10^{-8}$ ~\cite{Vesna2021}. So far these measurements have only been compared to nuclear models which put clustering dynamics in by hand. Calculations from first principles of parity violation in these reactions would be highly desirable,
in view of the scarce information on NN weak amplitudes.   

At the same time polarized neutron beams of slightly higher energies in the eV-keV range were used to conduct a broad survey of measurements of the very large parity-odd effects which can appear in p-wave compound neutron-nucleus resonances in heavy nuclei. The mechanism for the tremendous amplification of P-odd effects in neutron-nucleus resonances in heavy nuclei from resonance-resonance mixing, which can reach $10^{6}$ in the P-odd amplitude, is well-enough understood that the effects were predicted theoretically~\cite{Sushkov:1982fa, Bunakov:1982is} before they were observed experimentally~\cite{Alfimenkov1983}.  The scientific purpose of this work was to gain information on NN weak interaction amplitudes and to use this data as a new type of probe of nuclear chaos. The nearest-neighbor energy spacings and the distribution of the widths of the sharp, long-lived compound neutron-nucleus resonances one finds in heavy nuclei just above neutron separation energy are already known for decades to obey statistical  distributions consistent with the expectations of random matrix theory as applied to the nuclear Hamiltonian, an approach pioneered by Wigner, Dyson, and Mehta, noting~\cite{Dyson:1972tm,Mehta:randombook} and references therein. 
If one expresses the wave functions of these resonances in terms of their Fock space components, one expects in this view that the relative weights of the 
some $10^{6}$ or so independent components can be treated as random variables. These very complicated nuclear states are thought to be so complicated that robust statistical predictions for certain observables can be made. In particular, researchers extended the statistical theory to parity violation in heavy nuclei and made a prediction for $<M^{2}>$,  the mean square parity-odd mixing matrix element $M$ between s-wave and p-wave neutron-nucleus resonances, in terms of effective $\Delta I=0$ and $\Delta I=1$ NN weak amplitudes inside nuclei. Since the resonance-resonance mixing effects can in some cases amplify P-odd effects in these p-wave resonances from their generically-expected $10^{-7}$ size to sizes as large as $10^{-1}$ as noted above, the TRIPLE collaboration was able to measure 75 P-odd asymmetries above $3\sigma$ in statistical significance over a decade-long experimental campaign in several heavy nuclei in measurements of the parity-odd asymmetry $\Delta \sigma_{P}$ of spin-dependent transmission of longitudinally-polarized neutrons. The results were in qualitative agreement with the size of NN weak amplitudes expected from the dimensional analysis arguments presented above. 

Can the predictions of nuclear statistical spectroscopy for the variance of P-odd mixing matrix elements be placed on a more quantitative footing? The experimental data and the statistical analysis methods for $<M^{2}>$ exist: what remains is to measure the angular momentum quantum numbers of the resonances, determine the NN weak amplitudes and their modification in the medium of a heavy nucleus, and perform the statistical calculation in light of the new data on NN weak amplitudes. Updated estimates from a mean field approach~\cite{Flambaum:2022} report results for  $<M^{2}>$ in agreement with the TRIPLE data. The results from a statistical calculation in the nuclear shell model~\cite{Tomsovic:1999yb} used NN weak coupling inputs from DDH and are therefore well worth revisiting in light of the subsequent progress discussed above. It would be interesting to compare these results with those from the atomic parity violation experimental work described in the next section. 
A successful comparison between theory and experiment in this observable could also help quantify the results from future searches for time reversal violation in neutron-nucleus resonances reactions in polarized and tensor-aligned targets~\cite{Gudkov:1990tb, Bowman:2014fca, Fadeev:2019}.    
 
\subsection{Leptonic Probes of Hadronic Parity Violation: The Role of Atomic Parity Violation \label{sec:apv-anapole}}

Section \ref{sec:APV} 
laid out how 
atomic measurements of nuclear-spin-dependent (NSD) parity violation extract the combination $\kappa_{\rm total} = \kappa_a + \kappa_Z + \kappa_{\rm hf}$ (Eq. (\ref{eq:kappa_total})), along
with the theoretical calculations needed to 
isolated $\kappa_a$ from $\kappa_{\rm total}$
with the expectation that 
$\kappa_a$ should prove dominant in heavy nuclei. 
Here we discuss what these measurements have delivered, what the resulting ``anapole tension'' with other hadronic parity-violating 
observables 
may imply, 
and what the next generation of experiments aims to achieve.\\

\noindent \underline{The DDH Model and the ``Anapole Tension''}\\

To integrate atomic measurements of the anapole moment ($\kappa_a$) with broader hadronic physics constraints, $\kappa_a$ must be formalized within a 
framework of the weak NN interaction. The historical foundation for this is the phenomenological meson-exchange model of DDH~\cite{DDH1980} that has been 
much discussed in previous sections. 
Within this framework, the nuclear anapole moment of a given isotope is computed by evaluating the matrix elements of the anapole operator from Eq.~(\ref{as}) between the parity-admixed ground states obtained from Eq.~(\ref{gswf}), with the parity-violating NN potential built from the 
DDH couplings. 
The result is a numerical mapping from the DDH coupling space to
$\kappa_a$, supported 
by a nuclear-structure calculation that depends on many-body correlations, core polarization, and two-body exchange currents~\cite{Haxton2001,Haxton2002,Flambaum1997,Liu2007}.
In Sec.~\ref{sec:th-SM} we have 
discussed how the different 
constraints on the underlying DDH
couplings compare, 
and noted how 
the determined constraints on the
DDH couplings from the measurement 
of parity violation in $^{133}$Cs 
and the
extraction of 
its associated anapole 
moment~\cite{Wood1997,Haxton2001}, 
do 
not overlap with the corresponding bands from other hadronic parity-violating observables~\cite{Flambaum:1997um,Haxton2002,Safronova2018}. For example, the $^{18}$F radiative-decay analysis constrains $|h_\pi^1| < 1.1 \times 10^{-7}$ (at 67\% CL)~\cite{Haxton1981,Adelberger_beta2PV:1983zz}; the NPDGamma measurement~\cite{Blyth2018} yields $h_\pi^1 = (2.6 \pm 1.2_{\textrm{stat}}\pm0.2_{\textrm{sys}}) \times 10^{-7}$, 
which compares favorably to the \textit{ab initio} 
computation of $(2.14 \pm 0.21) \times 10^{-7}$ in pQCD and LQCD 
in Eq.~(\ref{hpihere}), as well as its determination in 
ChEFT~\cite{deVries:2015pza}. 
Reproducing the measured $\kappa_a$ in $^{133}$Cs while respecting these constraints would require $|h_\rho^0 + 0.7\, h_\omega^0|$ at values well outside the DDH reasonable range and inconsistent with all other hadronic PV observables at the $\sim3\sigma$ level. This noted 
mismatch is an 
expression of the long-standing ``anapole tension''~\cite{Haxton2001PRL}.

It is worth emphasizing that the $\kappa_a$ extraction, as well as
the size of the parity-violating 
atomic matrix element from 
which it is extracted, rely on 
inputs from atomic theory and 
experiment. This can impact the 
strength of the parity-violating
atomic matrix element, because it 
is the ratio of that quantity
to the parity-conserving
vector polarizibility $\beta$
that is measured~\cite{Wood1997}.
There have been disagreements in the outcomes
of different methods of its assessment in 
the needed $^{133}$Cs transition, and with 
advancements in theory and experiment they 
have been mitigated. For example, 
Quirk \textit{et al.}~\cite{Quirk2024}, with the use of 
new, precise theoretical values of the E1 transition 
moments~\cite{TranTanDev2023PhRvA.107d2809T}, 
have reported a high-precision measurement of the dc Stark shift of the $7s\, {}^2S_{1/2}$ level in $^{133}$Cs, yielding a value $4.7\sigma$ discrepant from the previous determination and providing a revised vector transition polarizability $\tilde\beta = 27.043(36)\, a_0^3$. This result {\it reduces}
a long-standing $\sim 0.7\%$ ($2.8\sigma$) discrepancy between competing $\beta$ determinations~\cite{Toh2019,TranTan2023} that had been a critical limitation in interpreting the Boulder measurement~\cite{Wood1997} in terms of a parity-violating atomic matrix element. With this result in 
hand, atomic theory is needed to extract 
$\kappa_{tot}$ --- here
relativistic coupled-cluster (RCC) recalculations of the $^{133}$Cs parity-violating transition amplitude~\cite{Chakraborty2024, Pandey2025} give
new values of $\kappa_{\rm tot}$, yielding, finally, different assessments of the anapole moments: 
$\kappa_a=0.119 (17)$~\cite{Pandey2025}, $\kappa_a=0.102 (16)$~\cite{Chakraborty2024}, to compare with the earlier results of 
$\kappa_a = 0.098 (16)$~\cite{Johnson2003} and $\kappa_a =0.090 \pm 0.016$~\cite{Haxton2002}. Assessments of the inputs that yield
these results and their implications are ongoing. 
This progress, although the situation is still evolving, 
reframes the experimental and theoretical priorities: an independent anapole measurement in a system with cleaner nuclear (and atomic) structure (e.g., light nuclei accessible via molecular systems, where ab initio no-core shell model calculations of the type discussed in Sec.~\ref{sec:many-body} are tractable) would test the theoretical framework 
directly, while improved IMSRG calculations of $\kappa_a$ in $^{133}$Cs itself~\cite{Belley2026} would test whether a residual discrepancy is a nuclear-structure artifact or a genuine signal of physics beyond the SM.

While the DDH framework remains the foundational historical benchmark for contextualizing the anapole tension, modern theoretical efforts have increasingly transitioned to assessments using 
ChEFT, as we have developed here. 
From a global perspective, ChEFT provides a controlled expansion and a consistent operator basis that can accommodate both few-body observables (e.g., neutron capture asymmetries) and many-body anapole extractions, albeit with different nuclear-structure 
inputs. Its structure is 
intrinsically connected to the underlying symmetries of Quantum Chromodynamics (QCD), and it gives 
a systematic expansion, moving the field beyond the phenomenological limitations 
of meson-exchange models. Please see Secs.~\ref{sec:intro} and~\ref{sec:theory} for a review of the state of the art.\\

\noindent \underline{The Experimental Landscape of Anapole Measurements}\\

Table~\ref{tab:NSD_landscape} summarizes the current experimental landscape for nuclear-spin-dependent atomic and molecular parity violation, organized by system class and indicating the dominant uncertainty in each case. Two features deserve emphasis at the outset. First, the existing heavy-atom measurements are not unanimous: the 1997 Boulder result in $^{133}$Cs~\cite{Wood1997} extracted a large nonzero $\kappa_a$, while the 1995 Seattle measurement in $^{205}$Tl~\cite{Vetter1995} yielded a result consistent with zero~\cite{Kozlov2002}, and the two constraints are mutually inconsistent within the DDH framework~\cite{Haxton2002}. Second, four broad strategies for new measurements are visible: (i) precision improvements on the heavy alkalis, in Cs, Fr; (ii) intrinsic enhancement of the PV mixing amplitude via small atomic energy denominators or advanved optical techniques, in Tl, Yb, Dy, and Sm; (iii) the diatomic molecular program, which exploits Zeeman-tunable rotational level crossings for amplification factors orders of magnitude beyond what is achievable in atoms; and (iv) the more recent polyatomic and radioactive-molecule programs, which target either calculable nuclear structure (light polyatomics, where ab initio NCSM calculations are tractable~\cite{Hao2020}) or maximal collective enhancement (octupole-deformed nuclei). We discuss each in turn, with emphasis on what each delivers for the anapole tension discussed above.

\begin{table*}[t]
\centering
\caption{Current and prospective experimental approaches to 
nuclear-spin-dependent atomic and molecular parity violation 
(NSD-APV). The ``enhancement'' column quotes the system-specific 
amplification factor for the NSD-PV signal as defined in the cited 
literature; these factors are not directly comparable across rows 
(see text). Status codes: M = measured; IP = in progress; 
P = proposed.}
\label{tab:NSD_landscape}
\renewcommand{\arraystretch}{1.3}
\begin{tabular}{@{}lllll@{}}
\toprule
System & Observable / Approach & Enhancement & Status & 
Dominant Uncertainty/Comment \\
\midrule
\multicolumn{5}{l}{\textit{Heavy Alkali}} \\
$^{133}$Cs 
  & Stark-interference, $6S \to 7S$~\cite{Wood1997} 
  & baseline 
& M~\cite{Wood1997}
  & Atomic theory; nuclear structure \\
$^{133}$Cs 
  & Coherent rf + Raman, direct $\kappa_a$~\cite{Damitz2024} 
  & --- 
  & P
  & Independent test of Boulder $\kappa_a$ \\
$^{210}$Fr 
  & Stark-interf., trapped atoms~\cite{Gwinner2022,Mariotti2014} 
  & $\sim 18\times$ Cs~\cite{Gomez2005}
  & IP
  & Trap systematics; isotope production \\
\midrule
\multicolumn{5}{l}{\textit{Heavy Atoms}} \\
$^{205}$Tl 
  & Optical rotation, $6P_{1/2} \to 6P_{3/2}$~\cite{Vetter1995} 
  & --- 
  & M~\cite{Vetter1995} 
  & Result consistent with zero~\cite{Kozlov2002} \\
$^{173}$Yb 
  & Stark + hyperfine, $^1S_0 \to {}^3D_1$~\cite{Antypas2019} 
  & $\sim100\times$ Cs~\cite{DeMille1995} 
  & IP
  & Many-body theory; hyperfine sys. \\
$^{163}$Dy 
  & Close opposite-parity levels~\cite{Nguyen1997} 
  & intrinsic, $\propto \tfrac{1}{\Delta E}$~\cite{Nguyen1997} 
  & IP 
  & Field control; level structure \\
$^{149}$Sm 
  & Optical rotation polarimetry~\cite{Lucas1998} 
  & intrinsic 
  & M~\cite{Lucas1998}
  & Atomic many-body theory (severe) \\
\midrule
\multicolumn{5}{l}{\textit{Diatomic Molecules}} \\
$^{137,138}$BaF
  & Zeeman-tuned rot. states, beam~\cite{DeMille2008} 
  & $\sim 10^{11}\times$ atoms~\cite{Altuntas2018} 
  & IP
  & Field inhomogeneities, state prep.\\
Various
  & Zeeman-tuned rot. states, ion trap~\cite{Karthein2024} 
  & $> 10^{12}\times$ atoms~\cite{Karthein2024} 
  & IP
  & Field inhomogeneities, state prep.\\
\midrule
\multicolumn{5}{l}{\textit{Polyatomic Molecules}} \\
Light triatomics 
  & Optical trap, $\ell$-doublets~\cite{Norrgard2019}
  & $\sim 70\times$ Cs~\cite{Norrgard2019} 
  & P
  & Demonstrated trapping~\cite{Hallas2023} \\
\midrule
\multicolumn{5}{l}{\textit{Octupole-Deformed Nuclei in Molecules}} \\
RaF, actinides
  & Various~\cite{GarciaRuiz2020,Karthein2024}
  & $\gg 10^3\times$ Cs~\cite{Flambaum1999} 
  & P 
  & Production rates; molec. spectroscopy \\
\bottomrule
\end{tabular}
\end{table*}

The 1997 Boulder Cs measurement~\cite{Wood1997} remains the only nonzero atomic NSD-PV result and the foundational constraint for modern anapole research, with improved Cs measurements underway \cite{Antypas2013, Damitz2024}. The FrPNC collaboration~\cite{Gwinner2022} aims to resolve the current tension between Cs and theory by pursuing precision Stark-interference measurements in trapped radioactive Fr isotopes, exploiting an $\sim 18\times$ PV signal enhancement over Cs from atomic structure scaling~\cite{Gomez2005}, building on the magneto-optical trapping of $^{210}$Fr~\cite{Simsarian1996}. Fr optical trapping was also demonstrated at INFN-LNL \cite{Mariotti2014}. A successful Fr measurement would clarify whether the Cs result is anomalous or whether the heavy-alkali set carries a common interpretive issue.

Preceding the Cs measurement, the Seattle Tl measurement~\cite{Vetter1995}, although consistent with zero~\cite{Kozlov2002}, provides an independent constraint that is not consistent with the Cs result within DDH. Three rare-earth and heavy systems exploit small intrinsic energy denominators or advanced optical techniques to amplify PV mixing. Ytterbium offers a large PV signal predicted by DeMille~\cite{DeMille1995} and refined by subsequent calculations~\cite{Porsev1995,Das1997,Dzuba2011}; while recent work~\cite{Antypas2019} has used isotopic comparisons to isolate the NSI weak charge, extracting the much smaller NSD signal requires hyperfine-resolved methodology and tight control of hyperfine-induced systematics~\cite{Tsigutkin2007,Kozlov2019}. Dysprosium remains the archetype of the near-degeneracy approach~\cite{Nguyen1997}, exploiting opposite-parity levels separated by a few cm$^{-1}$ or less. Samarium illustrates the limits of this strategy: the Oxford optical-rotation measurement~\cite{Lucas1998} reached impressive experimental sensitivity but could not be converted to a weak-charge constraint because of intractable atomic many-body theory. The Sm experience underscores that experimental sensitivity is necessary but not sufficient --- the atomic theory must be tractable for the result to constrain hadronic physics.

A major transition in the APV landscape is the shift toward heavy polar molecules, in which rotational energy levels of opposite parity can be tuned to near-degeneracy by external magnetic fields. The theoretical foundation for molecular parity violation was established by Kozlov, Labzovski \& Mitrushchenko ~\cite{Kozlov1991}, who first identified the field-tunable near-degeneracy mechanism. DeMille \textit{et al.}~\cite{DeMille2008} subsequently quantified enhancement factors of $10^3$ to $10^5$ relative to comparable atomic systems, depending on the specific molecule, transition, interrogation time, and applied field. ZOMBIES (DeMille group) uses Zeeman-tuned rotational levels in neutral BaF beams, with the $I=0$ isotope $^{138}$BaF serving as a methodological demonstration~\cite{Altuntas2018} ($\sim 10^{11}\times$ amplification over comparable atomic systems) before transitioning to the anapole-sensitive odd-isotope $^{137}$BaF. The NEPTUNE project pursues single trapped molecular ions in a Penning trap~\cite{Karthein2024} (first proposed by DeMille \textit{et al.}~\cite{DeMille2008}), trading throughput for exquisite control of state coherence and field environment. The ion trap approach provides a platform extensible to short-lived radioactive species, and thus, systematic scans of isotopes to decouple underlying parity-violating and nuclear effects.

Radioactive molecules containing octupole-deformed nuclei such as $^{225}$Ra exploit closely spaced parity doublets to amplify NSD-PV observables by orders of magnitude beyond the heavy-alkali baseline~\cite{Flambaum1999,Auerbach1996,Spevak1997}. The theoretical framework~\cite{IsaevHoekstraBerger2010} has been extended to laser-coolable polyatomic molecules~\cite{Isaev2016,Isaev2017}, and the experimental program achieved a major milestone with the first precision laser spectroscopy of short-lived RaF by Garcia Ruiz \textit{et al.}~\cite{GarciaRuiz2020}. Production and study of novel radioactive molecular candidates is now expanding rapidly across major radioactive-ion-beam facilities; we refer the reader to~\cite{Arrowsmith2024} for a comprehensive overview. The radioactive-molecule program targets the regime of maximal amplification and systematic scans across different isotopes to decouple the underlying parity-violating and nuclear effects.

Linear polyatomic molecules offer a different advantage for NSD-PV: closely spaced opposite-parity $\ell$-doublets are a generic feature of bending modes, allowing PV-sensitive level pairs to be tuned to degeneracy in magnetic fields roughly two orders of magnitude smaller than required in diatomics~\cite{Norrgard2019}. The proposal of Norrgard \textit{et al.}~\cite{Norrgard2019} targets $\sim 70\times$ improvement over current NSD-PV sensitivity using optically trapped light triatomics containing nuclei (Be, Mg, N, C) for which ab initio NCSM calculations of $\kappa_a$ are tractable~\cite{Hao2020}. Optical trapping of CaOH in $\ell$-doublet states~\cite{Hallas2023} have demonstrated the platform's viability. This program is distinct from the heavy-system measurements: a successful light-nucleus measurement, combined with the controlled theory, would isolate $\kappa_a$ from $\kappa_Z$ in a system where nuclear structure is not the dominant uncertainty, providing a clean test of the entire framework that the Cs and Tl measurements cannot.

While not the main focus of this article, it is important to note that parity-violating electron scattering (PVES) experiments are sensitive to NSD-PV effects \cite{Musolf1991}. A number of recent experiments — notably the backward-angle measurements of SAMPLE at MIT-Bates \cite{Ito2004}, G0 \cite{Androic2010} and SoLID \cite{Zheng2021, Arrington2023} at Jefferson Lab, and PVA4 at MAMI \cite{Baunack2009} — are directly sensitive to the effective neutral-current axial form factor $\widetilde{G}_A^{e,Z}$, which receives contributions from the nucleon anapole moment through electroweak radiative corrections \cite{Zhu2000}. Even in forward-angle measurements such as Qweak \cite{Androic2018}, NSD physics enters indirectly through the $\gamma Z$ box diagrams, whose evaluation requires knowledge of spin-dependent nucleon structure functions \cite{Gorchtein2011}. These PVES results provide constraints on hadronic parity violation --- in particular the PV pion-nucleon coupling $h_\pi^1$ --- that are highly complementary to those obtained from atomic PV measurements of nuclear anapole moments.
\section{Synergies\label{sec:syner}}

A primary goal of this research program is to 
characterize the low-energy NN weak interaction 
in a redundant way, to enable further tests of its nature. 
More experimental and theoretical work in atomic, molecular, and nuclear systems is needed to reach this goal. Continued extension of the NN weak EFT calculations to encompass more few-body systems is essential for the interpretation of measurements and is the subject of active ongoing work. 
Lattice gauge theory efforts to calculate NN weak amplitudes, as highlighted in Sec.~\ref{sec:lqcd}, 
may well succeed in computing 
the weak pion-nucleon coupling $h_\pi^1$, 
whereas the $\Delta I=2$ NN weak amplitude, 
which is computationally easier to access than the other NN weak amplitudes due to the absence of disconnected diagrams, continues to be an
aspiration. Additional work that could help 
develop 
insight into the hierarchy of isospin contributions 
to the low-energy constants of the PVTC NN potential
in ChEFT, as shown in Table \ref{tab:Ci},  would be most welcome. 

The main experimental approach available to conduct NN weak interaction experiments in few-body systems involves intense, low-energy beams of photons, neutrons, and protons interacting with few-nucleon nuclear targets, with either the beam or the target particles polarized to resolve parity-odd effects through pseudoscalar observables. As it is technically difficult to reverse the polarization of macroscopic ensembles of target nuclei quickly enough to suppress systematic errors at the required level of precision, in practice one needs polarized neutron, proton, or photon beams equipped with fast polarization reversal technology. Although the technology for fast reversal of polarized proton beams is well developed, to our knowledge no low energy intense polarized proton beams developed with proton parity violation measurements in mind are available. Polarized photon beams in the MeV energy range are in operation, and the technology to create them with high enough intensity to perform parity experiments is available, but at the moment no such beams are yet in existence.      

Prospects for additional measurements using low energy neutron beams are encouraging. Because it is possible to reverse the spin of low energy neutron beams with essentially no change in the phase space of the beam, so far all of the sensitive neutron parity violation measurements in few body systems are dominated by statistical errors, with systematic effects bounded to be about one order of magnitude smaller. This opens the way for improved measurements with higher intensity neutron beams. 

An apparatus in preparation for an improved n-$^{4}$He parity-odd neutron spin rotation experiment to measure the P-odd rotary power 
$d\phi / dz$ could reach a projected sensitivity of $10^{-8}$ rad/meter in a year of running on existing slow neutron beams at NIST, SNS, or ILL. This measurement can provide a strong constraint on a known linear combination of NN weak amplitudes. $d\phi / dz$ is expected to be $4.0 \times 10^{-7}$ rad/m based on the existing data on weak couplings combined with the most recent calculation~\cite{Lazauskas2019}, which implies that $d\phi / dz$ in $n+^{4}He$ is dominated by the isoscalar $\rho$ NN weak coupling $h^{0}_{\rho}$. It would be very worthwhile to perform an independent calculation of this observable using an alternate theoretical approach, such as Green's Function Monte Carlo extended to low energy  or Nuclear Lattice Effective Field Theory, that has
proven to work remarkably well for scattering, see e.g.~\cite{Elhatisari:2015iga}. 

Additional NN weak-interaction experiments could be conducted at  a future neutron beam at the European Spallation Source, which could produce a pulsed slow neutron beamline with time-averaged intensity comparable to the most intense existing reactor-based neutron sources. The value of the ESS for this physics lies in its unparalleled combination of high intensity slow neutron beams, needed to reach the statistical accuracy to see NN weak effects, combined with neutron energy information using neutron time of flight from this pulsed neutron source, which we know from experience is very important for the suppression of systematic errors for the ppb-level sensitivity to parity-odd asymmetries needed to see NN weak interaction effects in the presence of the strong interaction. Two examples of possible NN weak experiments which can take special advantage of the strengths of the ESS are (1) neutron-proton parity-odd spin rotation, which is one of the few experimentally-accessible observables with sensitivity to the $\Delta I=2$ NN weak amplitude, and (2) parity-odd gamma asymmetry in $\vec{n}+d \to t+\gamma$, which is a sufficiently simple system to be analyzed theoretically in terms of two-body NN weak amplitudes with high reliability. Another observable sensitive to the $\Delta I=2$ NN weak amplitude is parity-odd photodisintegration in $\vec{\gamma}+d \to n+p$ very near threshold, which could be pursued in principle with a sufficiently intense polarized photon beam, such as at an upgraded HiGS facility~\cite{Howell:2020nob}.

Turning to studies of 
nuclear spin-dependent parity violation in atoms and molecules, we anticipate
that we may well be able to measure 
trends in parity-violating 
observables with $Z$ and $A$ 
at sufficient precision 
to allow us to identify 
non-SM 
sources of parity violation
if they are present, 
or simply to further constrain 
long-popular extensions of the SM~\cite{Marciano1990}, 
without requiring precise theoretical assessments of
the associated SM processes.

\section{Summary} 

In this article we have developed how the 
hadronic-parity-violating effects observed in 
low-energy processes with nucleons and nuclei emerge from the combined dynamics of the weak and 
strong interactions of the SM.  
We have made explicit connection between 
the origins of this physics at the weak scale 
and the approximate physical scale of hadron dynamics, 
through QCD RG
techniques. 
We have described how chiral field theory 
translates the symmetries of the SM into 
a systematic framework in hadron degrees of 
freedom that is predictive once its low-energy constants
are determined, typically through experimental measurements. 
Through the work discussed in this article, however, 
we anticipate that we may ultimately be able to calculate 
these inputs in a well-controlled nonperturbative 
framework, such as LQCD --- thus giving us a completely
self-contained theoretical framework in which to 
frame the study of hadronic parity violation at low
energies. Within that setting, although still an 
aspiration, the theoretical frontier could focus on  
attaining theoretical control over hadronic parity
violation observables involving ever more massive
nuclei. Here the emergence of new computational 
frameworks such as nuclear lattice effective theory (NEFT)~\cite{Lahde:2019npb,Lee:2025req}, or the use of 
emulators to finesse the outcome of 
large-scale nuclear computations without
direct computation~\cite{Belley:2025nkn,Heihoff:2026ycq} (and references therein) --- may 
truly open new frontiers --- so that we may finally be able to return 
to where we started, to the 
study of parity violation in the compound nucleus, 
say --- and know the place for the very first time (with a nod to 
T.S. Eliot's ``Little Gidding'').

\begin{ack}[Acknowledgments]

SG and GM acknowledge partial support of this work from the U.S. Department of Energy, Office of Science, 
Office of Nuclear Physics under contract DE-FG02-96ER40989. 
JK acknowledges support from the Cyclotron Institute at Texas A\&M University.
The work of UGM was supported in part by the Deutsche Forschungsgemeinschaft (DFG, German Research Foundation) under Germany's
Excellence Strategy – EXC 3107 – Project-ID~533766364, by the European Research Council (ERC) under the European Union's Horizon 2020 research and innovation programme (EXOTIC, grant agreement No. 101018170) and by the CAS President's International Fellowship Initiative (PIFI) (Grant No.~2025PD0022).
PN acknowledges support from the Natural Sciences and Engineering Council of Canada (NSERC) Grant No. SAPIN-2022-00019. TRIUMF receives federal funding via a contribution agreement with the National Research Council of Canada.
WMS acknowledges support from US National Science Foundation grant PHY-2209481 and from the Indiana University Center for Spacetime Symmetries. 
\end{ack}

\seealso{article title article title}

\bibliographystyle{elsarticle-num}

\bibliography{reference.bib,hpv_review.bib}

\end{document}